# Formulations of the Elastodynamic Equations in Anisotropic and Multiphasic Porous Media from the Principle of Energy Conservation

Yinqiu Zhou[1,2,3], Xiumei Zhang[1,2,3], Lin Liu[1,2,3], Tingting Liu[1,3], and Xiuming Wang[1,2,3,*]

[1] State Key Laboratory of Acoustics, Institute of Acoustics, Chinese Academy of Sciences, Beijing 100190, People's Republic of China

[2] School of Physics Sciences in University of Chinese Academy of Sciences, Beijing 100149, People's Republic of China

[3] Beijing Engineering Research Center for Drilling and Exploration, Chinese Academy of Sciences, Beijing 100190, People's Republic of China

* **Email:** wangxm@mail.ioa.ac.cn

Elastodynamic equations have been formulated with either Newton's second law of motion, Lagrange's equation, or Hamilton's principle for over 150 years. In this work, contrary to classical continuum mechanics, a novel strategic methodology is proposed for formulating general mechanical equations using the principle of energy conservation. Firstly, based on Hamilton's principle, Hamilton's equations, Lagrange's equation, and the elastodynamic equation of motion are derived in arbitrarily anisotropic and multiphasic porous elastic media, for the first time. Secondly, these equations are all formulated using the principle of energy conservation for the related media. Both formulation results using the two kinds of principles are compared and validated by each other. The advantages of our methodology lie in that, the elastodynamic equation of motion, Lagrange's equation, and Hamilton's equations in continuum mechanics are directly formulated using a simple constraint of energy conservation without introducing variational concepts. It is easy to understand and has clear physical meanings. Our methodology unlocks the physics essences of Hamilton's principle in continuum mechanics, which is a consequence of the principle of energy conservation. Although the linear stress-strain constitutive relation is considered, our methodology can still be used in a nonlinear dynamical system. The methodology also paves an alternative way of treating other complex continuous dynamical systems in a broad sense. In addition, as an application, the continuity conditions at various medium interfaces are also revisited and extended using our proposed approach, which explains the law of reflections and refractions.

**Keywords:** Energy conservation, multiphasic porous media, Hamilton's principle, elastodynamic equations, boundary conditions.

## 1. Introduction

The elastodynamic behaviors in inhomogeneous anisotropic and fully elastic equivalent media have been treated thoroughly, and the elastodynamic equations of motion are essentially based on Newton's second law of motion [1−10] in homogenous equivalent elastic media, or Hamilton's principle [11−21] in various rather complex media. In the latter, all impressed forces of the monogenic type are absorbed in the usual way into Lagrange's density function, while the polygenic forces are given by their virtual work [11−14].

Porous media, composed of solid skeleton and pore fluids, such as underground reservoir media containing oil, gas, and water, permafrost in the Arctic region, quick-frozen foods, etc., exist widely in nature. Understanding wave propagation in such media for oil exploration, earthquake disaster prediction, hydrological environment monitoring, and frozen food monitoring is extremely important. The commonly used method to construct elastodynamic equations of motion for such complex media is Hamilton's principle, or the principle of least action, such as for piezoelectric materials [22], and porous media [23−33].

However, the achievements of a unified theory of multiphase continuum poroelasticity, capable of addressing multiphase systems with any range of compressibility of the constituents, still represent a challenge for theoretical and applied continuum mechanics, even for a simpler two-phase problem [33].

Biot paved the way to tackle poroelasticity problems in two-phase porous media filled with a single fluid [23−26]. He assumed that there was relative motion between pore fluid and solid phases, and this coupling was characterized by an apparent density to generate kinetic energy included in Lagrange's density function in the case of wavelength much larger than the pore space or grain sizes such that in any field point there would be fluid and solid components [23], which seemed unrealistic. Biot's assumptions of the coupling between fluid and solid phases were confirmed by Plona who observed the slow compressional waves that were theoretically predicted by Biot [34], which means Biot's theory could be used in describing wave propagation in two-phase porous media for low-frequency range to some extent.

Using Biot's theory, Pride et al. proposed a dual porous medium model to describe the propagation in a variety of complex rock structures [35]. Furthermore, Ba et al. derived Biot-Rayleigh's equation in a dual-porous medium, which is a combination of one fluid and two types of solid matrixes in dual-porous media. They generalized the theory to the special case of two fluids and one type of solid skeleton [32]. Bedford and Stern proposed the theory of liquid-saturated porous media containing bubbles based on the idea of mixture theory [36], which is consistent with Biot's theory except that the inertial effect caused by the bubble vibration was included. Tuncay and Corapcioglu used the volume average theory to investigate wave propagation in two immiscible Newtonian fluid-saturated poroelastic media [37]. Santos et al. established the wave theory of two viscous immiscible fluid-saturated porous media using the principle of compensated virtual work and Lagrange's approach [29,30]. Leclaire et al. derived momentum conservation equations for frozen porous media by using Lagrange's equation [38]. Furthermore, they studied the permafrost layer problems with one solid and one fluid in the pores. They predicted three types of longitudinal waves and two types of shear waves. Carcione and Seriani proposed the Carcione–Leclaire three-phase model, in which direct contact between solid particles and ice particles was considered [31]. Based on the uniform bubble distribution model, Carcione et al. studied the propagation of elastic waves in partially saturated porous media by using Biot's theory [39]. It was believed that the attenuation and dispersion of elastic waves were mainly caused by the energy transfer between different mode waves.

From the above reviews, it is seen that Biot's theory is generally believed to be semi-phenomenological, and more rigorous acoustic wave theories are based on the homogenization method [40] and the volume averaging method [41,42], which link the thermal microscopic and macroscopic situations. However, although the latter two methods are relatively rigorous, a lot of parameters are involved and it is difficult to obtain them experimentally. Thus, it is popular today that, dynamical equations of wave propagation in a porous medium filled with different matrixes and porous fluids are derived from Biot's theory framework based on Hamilton's principle or Lagrange's equation.

On the one hand, Hamilton's principle, or the principle of least action [43], although dominating the heart of physics and used in poroelasticity, is not explained well clearly in Lagrange's analytical mechanics, and this principle is presented in almost all textbooks beyond comprehension [12]. Recently, rather than using the existing principles, such as Hamilton's principle, Lagrange's equation, or Newton's second law of motion, a methodology for tackling dynamical problems in many-particle systems, is proposed, i.e., starting from the equation of conservation of energy, Lagrange's equation, Hamilton's equations, and Hamilton-Jacobi's equation are derived [44]. The work is also seen for an equivalent continuous system in identifying wave equations of motion in a piezoelectric medium [45,46].

On the other hand, most of the existing wave theories in porous media are limited to the case of two-phase three-component media [33,38] and the two-fluid and one-solid media [29,30]. In practice, multiphase and multicomponent media, such as acoustic monitoring for frozen foods, oil, gas, and cement material sandstones, and sandstone-bearing reservoir characterizations are encountered. Therefore, unified poroelasticity in an arbitrary anisotropic and multiphasic porous medium still left unresolved is important to be treated [33].

In this work, firstly we will give the formulations of the general mechanical equations in an arbitrarily anisotropic and multiphasic porous medium from Hamilton's principle and analyze the treatments of the boundary conditions. Furthermore, contrary to the existing Hamiltonian mechanics, we will formulate the general mechanical equations, including Lagrange's equation, Hamilton's equations, and elastodynamic equation of motion for the arbitrarily anisotropic and multiphasic porous medium under the framework of the principle of energy conservation. Finally, as an application example, we will revisit the stress and displacement continuity at various interfaces by using energy conservation constraints to explain Snell's law of reflection and refraction. Therefore, a different methodology from existing continuum mechanics will be proposed for formulating the general mechanical equations in an arbitrarily anisotropic and multiphasic porous medium

## 2. Physical model descriptions

### 2.1 *Assumptions*

The assumptions are stated in the following [38,48]. The displacement, strain, and particle velocity in an elastic medium are all small, so the Eulerian and Lagrangian formulations are consistent. The strain equation, dissipative force, kinetic energy, and momentum are linear (elastic strain energy, dissipative potential, and kinetic energy are of the quadratic type); the principles of continuum mechanics can be applied to measurable macroscopic values, while the macroscopic quantities are the volume averages of the corresponding microscopic quantities; the wavelength related to the macroscopic fundamental volume is large compared to the size of the base volume, that is, compared to the size of the basic volume, and this volume has well-defined properties such as porosity, permeability, and elastic modulus; the condition is adiabatic, and no heat exchange is considered in the process; the stress distribution in the fluid is hydrostatic; the liquid phase is continuous, while the matrix consists of solid and connected pores with disconnected pores make no contributions to the porosity.

2.2 *Model and variable descriptions*

Suppose that there are wave motions in an arbitrarily anisotropic and multiphasic porous medium. The number of material compositions making up the anisotropic porous medium is $M \geq 1$. For example, in gas hydrate-bearing formation, there are 3 compositions or components in the medium. They are water and gas-hydrate filled in the formation pores, while there is one kind of solid skeleton. We arbitrarily extract a volume $\Omega$ from this porous medium, the closed boundary surface is $\partial \Omega$. For Medium composition $\alpha$, let $u_i^{(\alpha)} = u_i^{(\alpha)}(x_i, t)$ be particle displacement components, and

$$\dot{u}_i^{(\alpha)} = \frac{du_i^{(\alpha)}(x_i,t)}{dt} = \frac{\partial u_i^{(\alpha)}(x_i,t)}{\partial t}, \tag{1a}$$

are the particle velocity components. The above equation is consistent with that in [8] for one component-bearing media. The cosine directional of the displacement is defined as $l_i$, $i = 1, 2, 3$. The elastic strain tensor components are defined as [38,48]

$$e_{ij}^{(\alpha)} = \frac{1}{2}(u_{i,j}^{(\alpha)} + u_{j,i}^{(\alpha)}), \tag{1b}$$

in Composition $\alpha$. It should be noted that our later discussions $\alpha = 1, 2, ..., M$, which means our treatments can be extended to an arbitrarily anisotropic and multiphasic porous medium. In the above strain formula [37,47],

$$u_{i,j}^{(\alpha)} = \frac{\partial u_i^{(\alpha)}(x_i,t)}{\partial x_j}, i, j = 1, 2, 3,$$

where $x_j$ is coordinate of $j$-direction for a Cartesian coordinate system. For example, when $M = 3$ for gas-hydrate bearing formation, the first medium is a solid skeleton; the second is a liquid (water) in the pores, and the third is a soft solid (gas-hydrate) in the pores. We denote $\rho_{mn}^a = \rho_{nm}^a$, a symmetric variable [48], as the apparent densities with Medium $m$ and Medium $n$ being coupled with each other; while $\rho_{11}^a, \rho_{22}^a, \rho_{33}^a$ are the densities of Media 1, 2, and 3, respectively. Note that the superscript $a$ in $\rho_{\alpha\beta}^a$ (rather than the Latin symbol $\alpha$) represents the apparent density. As the number of material compositions or components is larger than 3, we may denote the composition density of the medium as $\rho_{11}^a, \rho_{22}^a, \rho_{33}^a, ..., \rho_{\alpha\alpha}^a, ..., \rho_{MM}^a$. In this case, the number of medium compositions can be larger than 3, so that the coexisting compositions in the media can be any number as expected, i.e., $\alpha = 1, 2, ..., M$.

## 3. Formulations of general mechanical equations

Firstly, we start from Hamilton's variational principle, or simply, Hamilton's principle, and investigate the conventional mechanics to formulate Hamilton's canonical equations (or simply named Hamilton's equations), Lagrange's equation, and elastodynamic equation of motion in arbitrarily anisotropic and multiphasic porous media. Secondly, we formulate the existing or general mechanical equations from the principle of energy conservation in the arbitrarily anisotropic and multiphasic porous media.

The density functions of kinetic energy and elastic potential energy in a continuum with one composition are respectively defined as [6−8]:

$$K_\rho = \frac{1}{2}\rho \dot{u}_i \dot{u}_i, \tag{1c}$$

$$V_\rho = V_\rho(e_{ij}) = V_\rho(u_{i,j}). \tag{1d}$$

In our work, if the subscript and/or superscript are repeated, Einstein's summation convention will be conducted.

For multiphase or multicomponent porous media, such as permafrost and gas hydrate reservoirs, there are two-solid and one-liquid (three-component) media. According to our previous assumptions, a volume $\Omega$ is arbitrarily chosen in the continuous medium of interest with the boundary $\partial\Omega$. These two energy density functions are related to bulk density, particle vibration displacements, and velocities of the two-solid and one-liquid media and the equivalent deformations of each component.

Different from the treatment of the kinetic energy density function in previous work [48], we assume that there may be relatively coupled motions between the three components of the porous medium that generate kinetic energy. Therefore, the kinetic energy density function has a similar form as in each medium component. So that the equivalent kinetic energy density function for the whole porous medium, for example, with 3 compositions can be written as

$$2T_\rho = \rho_{11}^a \dot{u}_i^{(1)}\dot{u}_i^{(1)} + \rho_{22}^a \dot{u}_i^{(2)}\dot{u}_i^{(2)} + \rho_{33}^a \dot{u}_i^{(3)}\dot{u}_i^{(3)} + \rho_{12}^a (\dot{u}_i^{(1)} - \dot{u}_i^{(2)})(\dot{u}_i^{(1)} - \dot{u}_i^{(2)}) + \rho_{23}^a (\dot{u}_i^{(2)} - \dot{u}_i^{(3)})(\dot{u}_i^{(2)} - \dot{u}_i^{(3)}) + \rho_{31}^a (\dot{u}_i^{(3)} - \dot{u}_i^{(1)})(\dot{u}_i^{(3)} - \dot{u}_i^{(1)}), \tag{2a}$$

Rearranging Eq. (2a), we have

$$2T_\rho = \left(\rho_{11}^a + \rho_{12}^a + \rho_{13}^a\right)\dot{u}_i^{(1)}\dot{u}_i^{(1)} + \left(\rho_{21}^a + \rho_{22}^a + \rho_{23}^a\right)\dot{u}_i^{(2)}\dot{u}_i^{(2)} + \left(\rho_{31}^a + \rho_{32}^a + \rho_{33}^a\right)\dot{u}_i^{(3)}\dot{u}_i^{(3)} - 2\rho_{12}^a \dot{u}_i^{(1)}\dot{u}_i^{(2)} - 2\rho_{23}^a \dot{u}_i^{(2)}\dot{u}_i^{(3)} - 2\rho_{31}^a \dot{u}_i^{(3)}\dot{u}_i^{(1)}. \tag{2b}$$

Denote

$$\begin{cases} \rho_{11} = \rho_1 + \rho_{12}^a + \rho_{13}^a, \\ \rho_{22} = \rho_{21}^a + \rho_{22} + \rho_{23}^a, \\ \rho_{33} = \rho_{31}^a + \rho_{32}^a + \rho_{33}, \\ \rho_{12} = -\rho_{12}^a, \\ \rho_{23} = -\rho_{23}^a, \\ \rho_{31} = -\rho_{31}^a, \end{cases} \tag{2c}$$

By using $\rho_{mn} = \rho_{nm}, m, n = 1, 2, 3,$ and rearranging Eq. (2b), the kinetic energy density function can be simplified as

$$T_\rho = \frac{1}{2}\rho_{\alpha\beta}\dot{u}_i^{(\alpha)}\dot{u}_i^{(\beta)}, \tag{2d}$$

where $\alpha, \beta, i = 1, 2, 3$. By using this treatment, it is easy to obtain Eq. (2d) directly, as was usually done from the previous relations using the average value of macroscopic parameters and microscopic parameters [23–26,29–32,38,48]. In Eq. (2d), Einstein's summation convention was used.

Note that when the total number of the medium compositions is $M \geq 1$, the total kinetic energy density of multiphasic porous media can be briefly written as in Eq. (2d) with $i = 1, 2, 3,$ and $\alpha, \beta = 1, 2,..., M$. Therefore, Eq. (2d) can be easily extended to the case of arbitrarily anisotropic and multiphasic porous media. In the above formula, $\alpha, \beta$ are both the upper index and the lower index. If any index is repeated, it conforms to the Einstein summation convention. For example, in three-component anisotropic porous media, expanding the above formula, we get

$$2T_\rho = \rho_{11}\dot{u}_i^{(1)}\dot{u}_i^{(1)} + \rho_{12}\dot{u}_i^{(1)}\dot{u}_i^{(2)} + \rho_{13}\dot{u}_i^{(1)}\dot{u}_i^{(3)} + \rho_{21}\dot{u}_i^{(2)}\dot{u}_i^{(1)} + \rho_{22}\dot{u}_i^{(2)}\dot{u}_i^{(2)} + \rho_{23}\dot{u}_i^{(2)}\dot{u}_i^{(3)} + \rho_{31}\dot{u}_i^{(3)}\dot{u}_i^{(1)} + \rho_{32}\dot{u}_i^{(3)}\dot{u}_i^{(2)} + \rho_{33}\dot{u}_i^{(3)}\dot{u}_i^{(3)}. \tag{3a}$$

As is known, the apparent density between different Component $\alpha$ and Component $\beta$ is $\rho^a_{\alpha\beta} = \rho^a_{\beta\alpha}$. For any pair of medium components, it is easy to show from Eq.(2c), that

$$\rho_{\alpha\beta} = \rho_{\beta\alpha}, \tag{3b}$$

is held. Using the above variable symmetry, it turns out that Eq. (3a) is consistent with Eq. (2d).

Also, Eq. (2d) can be extended to arbitrarily anisotropic and multiphasic porous media, which means $\alpha, \beta \in [1, M]$, with $M$ could be equal to or larger than 1. In this case, for any number $M$, $\rho_{\alpha\beta}$ reads

$$\rho_{\alpha\alpha} = \rho^a_{\alpha 1} + \rho^a_{\alpha 2} + ... + \rho^a_{\alpha\alpha} + ... + \rho^a_{\alpha M}, \quad \text{and} \tag{3c}$$

$$\rho_{\alpha\beta} = -\rho^a_{\alpha\beta}, \alpha \neq \beta; \alpha, \beta = 1, 2, ..., M. \tag{3d}$$

In Eq. (3c), no Einstein's summation is used. Therefore, the kinetic energy density function shown in Eq.(2d) can be extended to arbitrarily anisotropic and multiphasic porous media together with particle velocity in Eq.(1a).

For a porous model, the equation of motion of porous media can be obtained from Lagrange's equation once the kinetic and potential energy densities are identified [21−30]. In our case, the potential energy density of an arbitrarily anisotropic and multiphasic porous medium can be expressed as [38]

$$V_\rho = V_\rho(e^{(\alpha)}_{ij}), i, j = 1, 2, 3, \text{and } \alpha = 1, 2, ..., M.$$

where $e^{(\alpha)}_{ij}$ is, as defined in Eq.(1b) before, the elastic strain components in Composition $\alpha$ for an arbitrarily anisotropic and multiphasic porous medium.

The density function of potential energy can be used to identify the stress-strain relationship or constitutive relationship of the medium of interest. Here we only consider the case where the constitutive relationship is linear, then the elastic coefficients are determined by using the strain energy density function $V_\rho$ [6,8].

Firstly, considering an anisotropic multiphasic porous medium with 3 compositions, we expand the elastic potential energy into Taylor's series concerning the strain components and keep only the second-order term for a linear relationship between the stress and strain, that is,

$$2V_\rho = 2V_\rho(0) + \frac{\partial^2 V_\rho}{\partial e^{(1)}_{ij} \partial e^{(1)}_{kl}} e^{(1)}_{ij} e^{(1)}_{kl} + \frac{\partial^2 V_\rho}{\partial e^{(2)}_{ij} \partial e^{(2)}_{kl}} e^{(2)}_{ij} e^{(2)}_{kl} + \frac{\partial^2 V_\rho}{\partial e^{(3)}_{ij} \partial e^{(3)}_{kl}} e^{(3)}_{ij} e^{(3)}_{kl} + 2\frac{\partial^2 V_\rho}{\partial e^{(1)}_{ij} \partial e^{(2)}_{kl}} e^{(1)}_{ij} e^{(2)}_{kl}$$

$$+ 2\frac{\partial^2 V_\rho}{\partial e^{(1)}_{ij} \partial e^{(3)}_{kl}} e^{(1)}_{ij} e^{(3)}_{kl} + 2\frac{\partial^2 V_\rho}{\partial e^{(3)}_{ij} \partial e^{(2)}_{kl}} e^{(3)}_{ij} e^{(2)}_{kl}. \tag{4a}$$

This treatment leads to linear stress-strain constitutive relations. However, our methodology is not limited to linear relations. For example, it can also be used in nonlinear fluid dynamical equation formulations [44]. The influence of the elastic potential energy on wave motion is about the variation of the space deformation. Therefore, we might as well take $V_\rho(0, 0, 0)$ to be zero for a reference value.

If the Einstein summation convention is used for the repeated upper and lower indices, the elastic potential energy may be briefly written as

$$2V_\rho = \frac{\partial^2 V_\rho}{\partial e^{(\alpha)}_{ij} \partial e^{(\beta)}_{kl}} e^{(\alpha)}_{ij} e^{(\beta)}_{kl}, \alpha, \beta = 1, 2, 3. \tag{4b}$$

To examine its validity, expanding Eq. (4b), yields

$$2V_\rho = \frac{\partial^2 V_\rho}{\partial e_{ij}^{(1)} \partial e_{kl}^{(1)}} e_{ij}^{(1)} e_{kl}^{(1)} + \frac{\partial^2 V_\rho}{\partial e_{ij}^{(1)} \partial e_{kl}^{(2)}} e_{ij}^{(1)} e_{kl}^{(2)} + \frac{\partial^2 V_\rho}{\partial e_{ij}^{(1)} \partial e_{kl}^{(3)}} e_{ij}^{(1)} e_{kl}^{(3)} + \frac{\partial^2 V_\rho}{\partial e_{ij}^{(2)} \partial e_{kl}^{(1)}} e_{ij}^{(2)} e_{kl}^{(1)} + \frac{\partial^2 V_\rho}{\partial e_{ij}^{(2)} \partial e_{kl}^{(2)}} e_{ij}^{(2)} e_{kl}^{(2)}$$

$$+ \frac{\partial^2 V_\rho}{\partial e_{ij}^{(2)} \partial e_{kl}^{(3)}} e_{ij}^{(2)} e_{kl}^{(3)} + \frac{\partial^2 V_\rho}{\partial e_{ij}^{(3)} \partial e_{kl}^{(1)}} e_{ij}^{(3)} e_{kl}^{(1)} + \frac{\partial^2 V_\rho}{\partial e_{ij}^{(3)} \partial e_{kl}^{(2)}} e_{ij}^{(3)} e_{kl}^{(2)} + \frac{\partial^2 V_\rho}{\partial e_{ij}^{(3)} \partial e_{kl}^{(3)}} e_{ij}^{(3)} e_{kl}^{(3)}.$$

(4c)

Considering the elastic potential energy density function is a continuous function of strain, the order of its second derivative concerning the strain component is commutative, i.e.,

$$\frac{\partial^2 V_\rho}{\partial e_{ij}^{(\alpha)} \partial e_{kl}^{(\beta)}} = \frac{\partial^2 V_\rho}{\partial e_{kl}^{(\beta)} \partial e_{ij}^{(\alpha)}}.$$

(4d)

Therefore, Eqs. (4a) and (4c) are identical.

As is discussed in Eq. (2d) for kinetic energy density function, Eq. (4b) can also be extended into an arbitrarily anisotropic and multiphasic porous medium, with $\alpha, \beta = 1, 2, ..., M$. In the following discussions, the physical quantities in Eq. (4d), denoted by

$$C_{ijkl}^{(\alpha,\beta)} = \frac{\partial^2 V_\rho}{\partial e_{ij}^{(\alpha)} \partial e_{kl}^{(\beta)}},$$

(5a)

will be seen as the elastic coefficients or elastic constants of the media, represented by the elastic energy density function. If $\alpha = \beta$, then it is the equivalent elastic constant of the medium; if the two are not equal, it characterizes the equivalent elastic constants when two media are coupled with each other.

From Eqs. (1b), (4d) and (5a), it is easy to prove that

$$C_{ijkl}^{(\alpha,\beta)} = C_{ijkl}^{(\beta,\alpha)} = C_{klij}^{(\alpha,\beta)} = C_{jilk}^{(\alpha,\beta)}.$$

(5b)

In the following derivations, the symmetries of elastic constants as shown in Eq. (5b) will be used later. By using Eqs. (4b), (5a) and (5b), the elastic potential energy density function for an arbitrarily anisotropic and multiphasic porous medium can be written as

$$V_\rho = \frac{1}{2} C_{ijkl}^{(\alpha,\beta)} e_{ij}^{(\alpha)} e_{kl}^{(\beta)}.$$

(6a)

The above equation degenerates to the one when the media only contain one composition, as shown in [13,14], that is,

$$V_\rho = \frac{1}{2} C_{ijkl}^{(1,1)} e_{ij}^{(1)} e_{kl}^{(1)}, \text{ or}$$

$$V_\rho = \frac{1}{2} C_{ijkl} e_{ij} e_{kl} = \frac{1}{2} C_{ijkl} u_{i,j} u_{k,l}.$$

(6b)

Also, since it is assumed that small deformations and the linear relationship between stress and strain are considered, the constitutive equations of stress and strain can be established using the elastic constants determined by elastic potential energy in Eq. (6a), namely,

$$\sigma_{ij}^{(\alpha)} = \frac{\partial V_\rho}{\partial e_{ij}^{(\alpha)}} = C_{ijkl}^{(\alpha,\beta)} e_{kl}^{(\beta)}.$$

By using Eqs. (1b) and (5b), it is easy to show that the above equation will become

$$\sigma_{ij}^{(\alpha)} = \frac{1}{2} C_{ijkl}^{(\alpha,\beta)} \left( u_{k,l}^{(\beta)} + u_{l,k}^{(\beta)} \right) = C_{ijkl}^{(\alpha,\beta)} u_{k,l}^{(\beta)},$$

(7)

which is defined as the generalized Hook's law in an arbitrarily anisotropic and multiphasic porous medium.

It should be noted that $\sigma_{ij}^{(\alpha)} = C_{ijkl}^{(\alpha,\beta)} e_{kl}^{(\beta)}$ is not the stress of Composition $\beta$ in the porous medium, but it is equivalent stress in Composition $\beta$. According to Eqs. (6a) and (7), we have

$$V_\rho = \frac{1}{2} \sigma_{ij}^{(\alpha)} e_{ij}^{(\alpha)}$$

The above Equation can also be extended into arbitrary medium compositions with $\alpha = 1, 2, ..., M; i, j = 1, 2, 3$. By using Eqs. (1b), (5b), and (7), it is easy to show that

$$V_\rho = \frac{1}{2} \sigma_{ij}^{(\alpha)} e_{ij}^{(\alpha)} = \frac{1}{2} \sigma_{ij}^{(\alpha)} u_{i,j}^{(\alpha)} = \frac{1}{2} C_{ijkl}^{(\alpha,\beta)} u_{i,j}^{(\alpha)} u_{k,l}^{(\beta)} \tag{8}$$

The kinetic energy and strain energy density functions in Eqs. (2d) and (8) respectively, can be used for arbitrarily anisotropic and multiphasic porous media, and they will be used for formulating general mechanical equations in the related media later.

### 3.1 General mechanical equations from Hamilton's principle

It is proclaimed that Hamilton's principle or Hamilton's variational principle is the general law of nature and perhaps the most fundamental law of nature revealed up to now [12], and it continues to hold its ground in the description of all the phenomena of nature [49,50]. Hamilton's principle and its variational variants become so important that they have constructed a founding stone of present-day theoretical physics [12].

It is no doubt that elastodynamics in porous media should also be under the umbrella of Hamilton's variational principle (Hamilton's principle) or its variants [33].

In the followings, starting from Hamilton's principle, we will formulate Hamilton's equations, Lagrange's equation, and the elastodynamic equation of motion in arbitrarily anisotropic and multiphasic porous media, and these formulated equations will be used for comparisons with those derived by our proposed methodology, which validates the derived equations with each other. In addition, we will clarify some ambiguous usages of boundary conditions during the equation formulations.

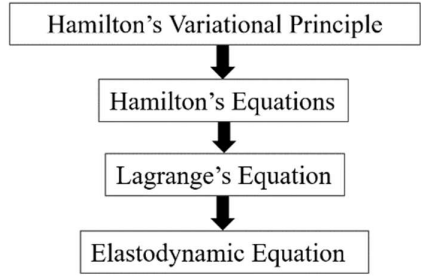

Fig. 1. The roadmap for formulating elastodynamic equations from Hamilton's principle.

The roadmap for formulating elastodynamics using Hamilton's principle is shown in Fig. 1. The procedure in formulating the related equations is that, from Hamilton's principle, Hamilton's equations will be derived, and then from Hamilton's equations, Lagrange's equations will be derived. Furthermore, the elastodynamic equation of motion will be derived from Lagrange's equation. This roadmap is the typical dynamical equation formulation procedures in classical continuum mechanics that have been developed since the establishment of Hamilton's principle.

#### 3.1.1 Hamilton's Canonical equations from Hamilton's principle

In this section we will extend Hamilton's equations for arbitrarily anisotropic and multiphasic porous media by using Hamilton's principle. Also, we will focus our attention on boundary treatments as there have been some ambiguous treatments on them. For example, the virtual displacements on the boundary are taken to be arbitrary [14,22,51], while they are also taken to be zero [6,13,52,53], which seems confusing to some extent.

Lagrange's density function is usually not time-explicit for a conservative system [4], so Hamilton's principle can be directly used. In this case, the Lagrange's and Hamilton's density functions are denoted by $L_\rho$ and $H_\rho$ respectively. For an elastic medium with one composition, Lagrange's density function is defined as [12,13]

$$L_\rho = L_\rho(u_i, \dot{u}_i, u_{i,j}) = T_\rho - V_\rho. \tag{9}$$

By using the relationship between the generalized momentum density function $\pi_i$, generalized velocity $\dot{u}_i$, the Hamilton density function through Legendre transformation is written as [12,13]

$$H_\rho = H_\rho(u_i, \pi_i, u_{i,j}) = \pi_i \dot{u}_i - L_\rho(u_i, \dot{u}_i, u_{i,j}), \tag{10a}$$

where the generalized momentum density is defined by [12,13]

$$\pi_i = \frac{\partial L_\rho}{\partial \dot{u}_i}. \tag{10b}$$

For a given volume in a conservative system, Hamilton's principle states that [12,51], one may use $\int_{t_1}^{t_2} \delta L dt = 0$ as a constraint to identify Lagrange's equation or the elastodynamic equation of motion, where Lagrange's function or the Lagrangian is obtained by $L = \iiint_\Omega L_\rho dv$. Therefore, for a given continuous medium, from

$$\int_{t_1}^{t_2} dt \, \delta \iiint_\Omega L_\rho dv = 0, \tag{11}$$

the correspondent Lagrange's equation and Hamilton's equations are determined with the assumed Lagrange's density function.

In continuum mechanics, the particle vibration displacement, velocity, and strain components are usually chosen as the independent generalized variables [6,25,32,38]. For any given anisotropic and multiphasic porous elastic medium, considering Legendre transformation shown in Eq. (10a), Lagrange's density function is written as

$$L_\rho = \dot{u}_i^{(\alpha)} \pi_i^{(\alpha)} - H_\rho, i = 1, 2, 3, \alpha = 1, 2, ..., M.$$

Lagrange's density function in Eq. (11) is represented by Hamilton's one, namely

$$\int_{t_0}^{t} dt \iiint_\Omega \delta(\dot{u}_i^{(\alpha)} \pi_i^{(\alpha)} - H_\rho) dv = 0. \tag{12a}$$

According to the variational principle, we have

$$\delta(\dot{u}_i^{(\alpha)} \pi_i^{(\alpha)} - H_\rho) = \dot{u}_i^{(\alpha)} \delta\pi_i^{(\alpha)} + \pi_i^{(\alpha)} \delta\dot{u}_i^{(\alpha)} - \frac{\partial H_\rho}{\partial u_i^{(\alpha)}} \delta u_i^{(\alpha)} - \frac{\partial H_\rho}{\partial \pi_i^{(\alpha)}} \delta\pi_i^{(\alpha)} - \frac{\partial H_\rho}{\partial u_{i,j}^{(\alpha)}} \delta u_{i,j}^{(\alpha)}. \tag{12b}$$

Therefore, Eq. (12a) can be written as

$$\int_{t_0}^{t} dt \iiint_\Omega [(\dot{u}_i^{(\alpha)} - \frac{\partial H_\rho}{\partial \pi_i^{(\alpha)}}) \delta\pi_i^{(\alpha)} + \pi_i^{(\alpha)} \delta\dot{u}_i^{(\alpha)} - \frac{\partial H_\rho}{\partial u_i^{(\alpha)}} \delta u_i^{(\alpha)} - \frac{\partial H_\rho}{\partial u_{i,j}^{(\alpha)}} \delta u_{i,j}^{(\alpha)}] dv = 0. \tag{13a}$$

In the above equation,

$$\int_{t_0}^{t} dt \iiint_\Omega \pi_i^{(\alpha)} \delta\dot{u}_i^{(\alpha)} dv = \iiint_\Omega (\pi_i^{(\alpha)} \delta u_i^{(\alpha)})\Big|_{t_0}^{t} dv - \int_{t_0}^{t} dt \iiint_\Omega \dot{\pi}_i^{(\alpha)} \delta u_i^{(\alpha)} dv = -\int_{t_0}^{t} dt \iiint_\Omega \dot{\pi}_i^{(\alpha)} \delta u_i^{(\alpha)} dv, \tag{13b}$$

where the constraints $\delta u_i^{(\alpha)}(t_0) = \delta u_i^{(\alpha)}(t) = 0$ have been used in Hamilton's principle. Furthermore, by using the following identity formula,

$$\frac{\partial H_\rho}{\partial u_{i,j}^{(\alpha)}} \delta u_{i,j}^{(\alpha)} = \left( \frac{\partial H_\rho}{\partial u_{i,j}^{(\alpha)}} \delta u_i^{(\alpha)} \right)_{,j} - \left( \frac{\partial H_\rho}{\partial u_{i,j}^{(\alpha)}} \right)_{,j} \delta u_i^{(\alpha)},$$

and substituting it into Eq. (13a), and considering Eq. (13b), we have

$$\int_{t_0}^{t} dt \iiint_\Omega \left\{ \left( \dot{u}_i^{(\alpha)} - \frac{\partial H_\rho}{\partial \pi_i^{(\alpha)}} \right) \delta \pi_i^{(\alpha)} - \dot{\pi}_i^{(\alpha)} \delta u_i^{(\alpha)} - \frac{\partial H_\rho}{\partial u_i^{(\alpha)}} \delta u_i^{(\alpha)} - \left( \frac{\partial H_\rho}{\partial u_{i,j}^{(\alpha)}} \delta u_i^{(\alpha)} \right)_{,j} + \left( \frac{\partial H_\rho}{\partial u_{i,j}^{(\alpha)}} \right)_{,j} \delta u_i^{(\alpha)} \right\} dv = 0. \qquad (13c)$$

By using Gauss's theorem [14],

$$\iiint_\Omega \left( \frac{\partial H_\rho}{\partial u_{i,j}^{(\alpha)}} \delta u_i^{(\alpha)} \right)_{,j} dv = \oiint_{\partial \Omega} \frac{\partial H_\rho}{\partial u_{i,j}^{(\alpha)}} l_j \delta u_i^{(\alpha)} ds$$

Eq. (13c) becomes

$$\int_{t_0}^{t} dt \iiint_\Omega \left\{ \left( \dot{u}_i^{(\alpha)} - \frac{\partial H_\rho}{\partial \pi_i^{(\alpha)}} \right) \delta \pi_i^{(\alpha)} - \left[ \dot{\pi}_i^{(\alpha)} + \frac{\partial H_\rho}{\partial u_i^{(\alpha)}} - \left( \frac{\partial H_\rho}{\partial u_{i,j}^{(\alpha)}} \right)_{,j} \right] \delta u_i^{(\alpha)} \right\} dv - \int_{t_0}^{t} dt \oiint_{\partial \Omega} \left( \frac{\partial H_\rho}{\partial u_{i,j}^{(\alpha)}} l_j \right) \delta u_i^{(\alpha)} ds = 0. \qquad (13d)$$

For a conservative system, there should be no energy exchange through the surface $\partial \Omega$. Therefore, by taking $\delta u_i^{(\alpha)} = 0$ on the surface $\partial \Omega$, Equation (13d) can be simplified into

$$\int_{t_0}^{t} dt \iiint_\Omega \left\{ \left( \dot{u}_i^{(\alpha)} - \frac{\partial H_\rho}{\partial \pi_i^{(\alpha)}} \right) \delta \pi_i^{(\alpha)} - \left[ \dot{\pi}_i^{(\alpha)} + \frac{\partial H_\rho}{\partial u_i^{(\alpha)}} - \left( \frac{\partial H_\rho}{\partial u_{i,j}^{(\alpha)}} \right)_{,j} \right] \delta u_i^{(\alpha)} \right\} dv = 0. \qquad (13e)$$

Apparently, $\delta u_i^{(\alpha)}$ and $\delta \pi_i^{(\alpha)}$ are arbitrary and independent of each other in the volume domain of the time interval. Let $\delta u_i^{(\alpha)}$ be zero on the enclosed surface $\partial \Omega$, By using the fundamental lemma proved by Gurtin [14, 54], the necessary conditions for Eq. (13e) to be held are

$$\begin{cases} \dot{u}_i^{(\alpha)} = \dfrac{\partial H_\rho}{\partial \pi_i^{(\alpha)}}, \\ \dot{\pi}_i^{(\alpha)} = -\dfrac{\partial H_\rho}{\partial u_i^{(\alpha)}} + \left( \dfrac{\partial H_\rho}{\partial u_{i,j}^{(\alpha)}} \right)_{,j}. \end{cases} \qquad (13f)$$

The above equations are the Hamilton's equations in an arbitrarily anisotropic and multiphasic porous medium. If there is only one material composition for the media of interest, i.e., $M = 1$, it will degenerate into Eqs. (16.30) and (16.31) given by Cline [14]. Later, Eq. group (13f) will be also validated with those derived from our methodology based on energy conservation equation.

By using Reynold's transport theorem [55,56], considering, we have

$$dH = \iiint_\Omega \left( \frac{\partial H_\rho}{\partial u_i^{(\alpha)}} du_i^{(\alpha)} + \frac{\partial H_\rho}{\partial \pi_i^{(\alpha)}} d\pi_i^{(\alpha)} + \frac{\partial H_\rho}{\partial u_{i,j}^{(\alpha)}} du_{i,j}^{(\alpha)} \right) dv. \qquad (14a)$$

On the other hand, using Eq. (10a) for an arbitrarily anisotropic and multiphasic porous medium, we have

$$H_\rho = H_\rho \left( u_i^{(\alpha)}, \pi_i^{(\alpha)}, u_{i,j}^{(\alpha)} \right) = \pi_i^{(\alpha)} \dot{u}_i^{(\alpha)} - L_\rho \left( u_i^{(\alpha)}, \dot{u}_i^{(\alpha)}, u_{i,j}^{(\alpha)} \right).$$

Therefore,

$$dH = d\iiint_\Omega H_\rho dv = d\iiint_\Omega \left[ \pi_i^{(\alpha)} \dot{u}_i^{(\alpha)} - L_\rho\left(u_i^{(\alpha)}, \dot{u}_i^{(\alpha)}, u_{i,j}^{(\alpha)}\right) \right] dv$$
$$= \iiint_\Omega \left( \pi_i^{(\alpha)} d\dot{u}_i^{(\alpha)} + \dot{u}_i^{(\alpha)} d\pi_i^{(\alpha)} - \frac{\partial L_\rho}{\partial u_i^{(\alpha)}} du_i^{(\alpha)} - \frac{\partial L_\rho}{\partial \dot{u}_i^{(\alpha)}} d\dot{u}_i^{(\alpha)} - \frac{\partial L_\rho}{\partial u_{i,j}^{(\alpha)}} du_{i,j}^{(\alpha)} \right) dv. \quad (14b)$$

From Eq. (10b), we know that the definition of generalized momentum $\pi_i^{(\alpha)}$ in the arbitrarily anisotropic and multiphasic porous media is

$$\pi_i^{(\alpha)} = \frac{\partial L_\rho}{\partial \dot{u}_i^{(\alpha)}}.$$

Substituting the above equation into Eq. (14b), yields

$$dH = \iiint_\Omega \left( \dot{u}_i^{(\alpha)} d\pi_i^{(\alpha)} - \frac{\partial L_\rho}{\partial u_i^{(\alpha)}} du_i^{(\alpha)} - \frac{\partial L_\rho}{\partial u_{i,j}^{(\alpha)}} du_{i,j}^{(\alpha)} \right) dv.$$

Comparing the above equation with (14a), yields

$$\iiint_\Omega \left[ \left( \frac{\partial H_\rho}{\partial u_i^{(\alpha)}} + \frac{\partial L_\rho}{\partial u_i^{(\alpha)}} \right) du_i^{(\alpha)} - \left( \dot{u}_i^{(\alpha)} - \frac{\partial H_\rho}{\partial \pi_i^{(\alpha)}} \right) d\pi_i^{(\alpha)} + \left( \frac{\partial L_\rho}{\partial u_{i,j}^{(\alpha)}} + \frac{\partial H_\rho}{\partial u_{i,j}^{(\alpha)}} \right) du_{i,j}^{(\alpha)} \right] dv = 0. \quad (14c)$$

Considering the first equation in Eq. group (13f), again, it is easy to show that arbitrary choice of integral volume $\Omega$ in Eq. (14c) leads to [14,54]

$$\left( \frac{\partial H_\rho}{\partial u_i^{(\alpha)}} + \frac{\partial L_\rho}{\partial u_i^{(\alpha)}} \right) du_i^{(\alpha)} + \left( \frac{\partial L_\rho}{\partial u_{i,j}^{(\alpha)}} + \frac{\partial H_\rho}{\partial u_{i,j}^{(\alpha)}} \right) du_{i,j}^{(\alpha)} = 0. \quad (14d)$$

Furthermore, since the variables $du_i^{(\alpha)}$ and $du_{i,j}^{(\alpha)}$ are also independent, and can be taken as arbitrary, followed by Arnold's procedure [57], Eq. (14d) leads to their coefficients of the independent variables $du_i^{(\alpha)}$ and $du_{i,j}^{(\alpha)}$ being zero, respectively, i.e.,

$$\begin{cases} \dfrac{\partial L_\rho}{\partial u_i^{(\alpha)}} = -\dfrac{\partial H_\rho}{\partial u_i^{(\alpha)}}, \\ \dfrac{\partial L_\rho}{\partial u_{i,j}^{(\alpha)}} = -\dfrac{\partial H_\rho}{\partial u_{i,j}^{(\alpha)}}. \end{cases} \quad (14e)$$

The above equations must be held during Legendre transformation, which will be used later for deriving Lagrange's equation from Hamilton's equations.

Eqs. (13f) are derived with Hamilton's principle in an arbitrarily anisotropic and multiphasic porous medium which is a continuous conservative system. It is guaranteed from the surface boundary conditions with $\delta u_i^{(\alpha)} = 0$ on $\partial\Omega$, which means there is no energy exchange through $\partial\Omega$. However, it is noted that, the choices of

$$\left.\delta u_i^{(\alpha)}\right|_{\partial\Omega} = 0, \quad (15)$$

make it impossible to mount the stress boundary conditions on $\partial\Omega$ in Eq. (13d) because whatever stress boundary conditions are, Eq. (15) makes all the stress boundary effects vanish. Therefore, this treatment should be avoided if there are stress-loaded boundary conditions. These constraints are naturally equivalent to traction-free boundary conditions as will be seen that

$$\sigma_{ij}^{(\alpha)} = \frac{\partial H_\rho}{\partial u_{i,j}^{(\alpha)}},$$

and $\sigma_{ij}^{(\alpha)} l_j$ is the normal stress components on the surface of $\partial\Omega$ for equivalent Composition $\alpha$ of the medium.

### 3.1.2 *Lagrange's equations from Hamilton's equations*

Lagrange's equation can be directly deduced from Hamilton's equations in Eqs. (13f) and (14e) for an arbitrarily anisotropic and multiphasic porous medium.

Substituting the first and second equations in Eq. group (14e) into the second equation in Eq. group (13f) which are Hamilton's equations, and considering the definition of generalized momentum in Eq. (10b), yields

$$\frac{d}{dt}\left(\frac{\partial L_\rho}{\partial \dot{u}_i^{(\alpha)}}\right) - \frac{\partial L_\rho}{\partial u_i^{(\alpha)}} + \left(\frac{\partial L_\rho}{\partial u_{i,j}^{(\alpha)}}\right)_{,j} = 0, \alpha = 1, 2, ..., M; i, j = 1, 2, 3, \tag{16}$$

which is Lagrange's equation, and it will degenerate into Lagrange's equation for three-phase porous media [31,39]. Therefore, Eq. (16) is a general Lagrange's equation for an arbitrarily anisotropic and multiphasic porous medium as a continuous conservative system.

### 3.1.3 *Elastodynamic equation of motion from Lagrange's equation*

According to the Lagrange density function in the arbitrarily anisotropic and multiphasic porous medium, the generalized momentum density is obtained and shown as

$$\pi_i^{(\alpha)} = \frac{\partial L_\rho}{\partial \dot{u}_i^{(\alpha)}} = \frac{\partial T_\rho}{\partial \dot{u}_i^{(\alpha)}}. \tag{17a}$$

Substituting Eq. (2d) into Eq. (17a), yields

$$\pi_j^{(m)} = \frac{1}{2} \frac{\partial\left(\rho_{\alpha\beta} \dot{u}_i^{(\alpha)} \dot{u}_i^{(\beta)}\right)}{\partial \dot{u}_j^{(m)}} = \frac{1}{2} \rho_{m\beta} \dot{u}_j^{(\beta)} + \frac{1}{2} \rho_{\alpha m} \dot{u}_j^{(\alpha)}.$$

By using $\rho_{\alpha\beta} = \rho_{\alpha\beta}$, further simplification of the above equation, yields

$$\pi_j^{(m)} = \rho_{m\alpha} \dot{u}_j^{(\alpha)}, \tag{17b}$$

where $i, j = 1, 2, 3$, and $m, \alpha = 1, 2, ..., M$. The above equation is expressed as the component of the equivalent generalized momentum density function in the $j$-direction for Component $m$ of an arbitrarily anisotropic and multiphasic porous medium. As mentioned above, the upper and lower indices in the above formula are repeated, and the Einstein summation convention for a two-solid and one-fluid anisotropic porous medium is followed

$$\begin{cases} \pi_j^{(1)} = \rho_{11} \dot{u}_j^{(1)} + \rho_{12} \dot{u}_j^{(2)} + \rho_{13} \dot{u}_j^{(3)}, \\ \pi_j^{(2)} = \rho_{21} \dot{u}_j^{(1)} + \rho_{22} \dot{u}_j^{(2)} + \rho_{23} \dot{u}_j^{(3)}, \\ \pi_j^{(3)} = \rho_{31} \dot{u}_j^{(1)} + \rho_{32} \dot{u}_j^{(2)} + \rho_{33} \dot{u}_j^{(3)}. \end{cases}$$

From Eq. (17b), since $\rho_{m\alpha}$ is independent of time, the derivatives of the generalized momentum density function with time for an arbitrary composition medium can be written as

$$\dot{\pi}_i^{(\alpha)} = \rho_{\alpha\beta} \ddot{u}_i^{(\beta)}. \tag{18}$$

Since the Lagrange density function is not the function of generalized displacement, that is,

$$\frac{\partial L_\rho}{\partial u_i^{(\alpha)}} = 0, i = 1, 2, 3; \alpha = 1, 2, ..., M.$$

Therefore,

$$\left(\frac{\partial L_\rho}{\partial u_{i,j}^{(\alpha)}}\right)_{,j} = -\left(\frac{\partial V_\rho}{\partial u_{i,j}^{(\alpha)}}\right)_{,j} = -\sigma_{ij,j}^{(\alpha)}. \tag{19}$$

Substituting Eqs. (18) and (19) into Eq. (16), yields

$$\rho_{\alpha\beta}\ddot{u}_i^{(\beta)} - \sigma_{ij,j}^{(\alpha)} = 0. \tag{20a}$$

The above equation is the elastodynamic equation of motion expressed in terms of displacement and stress. As mentioned before, the $\alpha$ represents the number of components or compositions that make up the multiphasic porous medium, which is not limited to 3 and can be extended to any number of compositions of the media. Therefore, Eq. (20a) is the elastodynamic equation of motion for an arbitrarily anisotropic and multiphasic porous medium.

As an example, for a two-solid and one-fluid (three-component) porous medium, unfolding Eq. (20a), we have

$$\begin{cases} \sigma_{ij,j}^{(1)} - \rho_{11}\ddot{u}_i^{(1)} - \rho_{12}\ddot{u}_i^{(2)} - \rho_{13}\ddot{u}_i^{(3)} = 0, \\ \sigma_{ij,j}^{(2)} - \rho_{21}\ddot{u}_i^{(1)} - \rho_{22}\ddot{u}_i^{(2)} - \rho_{23}\ddot{u}_i^{(3)} = 0, \\ \sigma_{ij,j}^{(3)} - \rho_{31}\ddot{u}_i^{(1)} - \rho_{32}\ddot{u}_i^{(2)} - \rho_{33}\ddot{u}_i^{(3)} = 0. \end{cases} \tag{20b}$$

If the acceleration of the above equation is rewritten as the derivative of the velocity with time, then the above equation is the first-order velocity-stress equation of motion under the conservative system, which is consistent with Carcione and Liu's results [31,58].

### 3.2 *General mechanical equations from the principle of energy conservation*

In all published works in continuum mechanics, it is claimed that various energy conservation formulae are 'derived' from Newton's second law of motion, Lagrange's equation, or Hamilton's principle. Even in quantum mechanics, Ehrenfest's theorem, related to energy conservation is 'deduced' by using Schrödinger's equation [59]. However, it is believed that the principle of conservation of energy is a law governing all the natural phenomena known to date and there is no known exception to this law [60]. Unfortunately, the principle of energy conservation is not like Hamilton's principle and its variants, such that although being lack of concrete physical meanings and having been questioned in serval aspects [61,62], it is the heart of theoretical physics, and is still dominating 21st century mostly in classical mechanics [63,64], fluid mechanics [65,66], elastodynamics [67], electrodynamics [68] and in quantum mechanics [68–74], etc.

In this section, totally different from the conventional treatments shown in 3.1, we will directly formulate Hamilton's equations, Lagrange's equation, and elastodynamic equation of motion by using the principle of energy conservation as an axiom, with the even less conditions than those used in Hamilton's principle, and then we will compare the obtained equations with those given by using Hamilton's principle. Furthermore, we will formulate the elastodynamic equations of motion in arbitrarily anisotropic and multiphasic porous media, i.e., we will establish the unified dynamical equation formulations of poroelasticity for arbitrarily anisotropic and multiphasic porous media based on Biot's model assumptions.

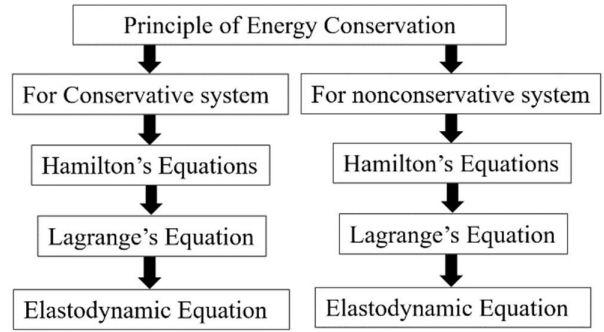

Fig. 2. Roadmap for formulating elastodynamic equations from the principle of energy conservation.

The roadmap for constructing elastodynamic equations based on the principle of energy conservation is shown in Fig. 2.

During our formulations of the general mechanical equations under the umbrella of the principle of energy conservation, two continuous systems are considered, i.e., the conservative and nonconservative systems. The formulation for the conservative system is used to compare with the results given by Hamilton's principle in validation. Nevertheless, it does not make a difference whether the system is conservative and nonconservative for the equation formulations from the perspective of energy conservation.

The axiom we use to derive general elastodynamic equations is the principle of energy conservation, which is obtained from the first law of thermodynamics by neglecting heat and chemical effects [44, 52]. The axiom is, the dynamical field of a continuous system has arbitrary possible states described by independent displacement, velocity, and strain components at the space and time domain, while the real dynamical field must abide by the principle of energy conservation, i.e., whatever dynamical field is presumed, the real physically dynamical field of the system must abide by the energy conservation and can be mathematically described as in the followings.

**Axiom.** For any physically dynamical field of a continuous system, the rate of its total energy of the system with time is equal to the rate of work done by external forces to the system with time.

From this simple energy conservation equation, we will formulate all general mechanical equations for an arbitrarily anisotropic and multiphasic porous medium. If the total mechanical energy is $E$, and the work done by external forces applied to the system is $W$, the axiom can be quantified as

$$\frac{dE}{dt} = \frac{dW}{dt} \tag{21}$$

in a multiparticle mechanical system [52,75], or

$$\frac{d}{dt}\left(\iiint_\Omega E_\rho dv\right) = \iiint_\Omega f_i^{(\beta)} \dot{u}_i^{(\beta)} dv + \oiint_{\partial\Omega} (\sigma_{ij}^{(\beta)} l_j \dot{u}_i^{(\beta)}) ds, \tag{22a}$$

in an arbitrarily anisotropic and multiphasic porous medium system with the number of $M$ compositions, where $E_\rho$ is the total mechanical energy density function. $f_i^{(\beta)}$ are the body forces, and $\sigma_i^{(\beta)} = \sigma_{ij}^{(\beta)} l_j$ are the stress components of Composition $\beta$ on the enclosed surface $\partial\Omega$ of an arbitrarily chosen volume $\Omega$. If $M = 1$, then Eq. (22a) degenerates into the case for one material component of the medium, and is consistent with, for example, those given by Achenbach and Miklowitz without body forces [6,8]. The term on the left-hand side of Eq. (22a) is the rate of the total mechanical energy with time in the volume $\Omega$ of interest, and those on the right-hand side are the work rate with time done by external body force into the volume and stress on the surface $\partial\Omega$ of the volume $\Omega$. If there are any other energy or work engaged in the system, they should be considered accordingly to satisfy Eq. (21) or Eq. (22a).

Since the Hamilton density function $H$ in the system is not an explicit function of time in a conservative system, it can be written as the sum of kinetic energy and potential energy densities [9,12,13]. Therefore,

$$\frac{dH}{dt} = \frac{dW}{dt}. \tag{22b}$$

In this case, Eq. (22b) in the conservative system is written as

$$\frac{dH}{dt} = \frac{d}{dt}\left(\iiint_\Omega E_\rho dv\right) = 0. \tag{22c}$$

However, it is noted that if $\partial H_\rho / \partial t \neq 0$, the system is nonconservative, and Eq. (22b) will not be held. In this situation, one still can link the total energy density with the Hamilton density function through the Lagrange density function [44]. Also, we will discuss the equation formulations in Sec. 3.2.4, for a nonconservative continuous system.

### 3.2.1 *Hamilton's equations from energy conservation for a conservative system*

In this sub-section, we will start from the law of conservation of energy and deduce Hamilton's equations in a continuous conservative system.

Using Eqs. (2d) and (8), the Hamilton density function can be expressed as

$$H_\rho = \frac{1}{2}\rho_{\alpha\beta}\dot{u}_i^{(\alpha)}\dot{u}_i^{(\beta)} + \frac{1}{2}C_{ijkl}^{(\alpha,\beta)}u_{i,j}^{(\alpha)}u_{k,l}^{(\beta)}. \tag{23}$$

According to Reynold's transport theorem [31], neglecting higher-order variable contributions, the first term on the left-hand side of Eq. (22a) can be written as

$$\frac{dH}{dt} = \iiint_\Omega \left( \frac{\partial H_\rho}{\partial u_i^{(\alpha)}}\dot{u}_i^{(\alpha)} + \frac{\partial H_\rho}{\partial \pi_i^{(\alpha)}}\dot{\pi}_i^{(\alpha)} + \frac{\partial H_\rho}{\partial u_{i,j}^{(\alpha)}}\dot{u}_{i,j}^{(\alpha)} + \frac{\partial H_\rho}{\partial t} \right) dv. \tag{24a}$$

On the other hand, using the Legendre transformation shown in Eq. (10a), the Hamilton density function is represented by the Lagrange density function. Therefore,

$$\begin{aligned} dH/dt &= \frac{d}{dt}\iiint_\Omega \left(\pi_i^{(\alpha)}\dot{u}_i^{(\alpha)} - L_\rho\right) dv \\ &= \iiint_\Omega \left( (\pi_i^{(\alpha)}\ddot{u}_i^{(\alpha)} + \dot{u}_i^{(\alpha)}\dot{\pi}_i^{(\alpha)} - \frac{\partial L_\rho}{\partial u_i^{(\alpha)}}\dot{u}_i^{(\alpha)} - \frac{\partial L_\rho}{\partial \dot{u}_i^{(\alpha)}}\ddot{u}_i^{(\alpha)} - \frac{\partial L_\rho}{\partial u_{i,j}^{(\alpha)}}\dot{u}_{i,j}^{(\alpha)} - \frac{\partial L_\rho}{\partial t} \right) dv. \end{aligned} \tag{24b}$$

According to the definition of generalized momentum density shown in Eq. (10b), Eq. (24b) can be simplified as

$$dH/dt = \iiint_\Omega \left( \dot{u}_i^{(\alpha)}\dot{\pi}_i^{(\alpha)} - \frac{\partial L_\rho}{\partial u_i^{(\alpha)}}\dot{u}_i^{(\alpha)} - \frac{\partial L_\rho}{\partial u_{i,j}^{(\alpha)}}\dot{u}_{i,j}^{(\alpha)} - \frac{\partial L_\rho}{\partial t} \right) dv. \tag{24c}$$

Subtracting Eq. (24a) by Eq. (24c) and rearranging them, yields

$$\iiint_\Omega \left[ \left(\frac{\partial H_\rho}{\partial u_i^{(\alpha)}} + \frac{\partial L_\rho}{\partial u_i^{(\alpha)}}\right)\dot{u}_i^{(\alpha)} + \left(\frac{\partial H_\rho}{\partial \pi_i^{(\alpha)}} - \dot{u}_i^{(\alpha)}\right)\dot{\pi}_i^{(\alpha)} + \left(\frac{\partial H_\rho}{\partial u_{i,j}^{(\alpha)}} + \frac{\partial L_\rho}{\partial u_{i,j}^{(\alpha)}}\right)\dot{u}_{i,j}^{(\alpha)} + \frac{\partial H_\rho}{\partial t} + \frac{\partial L_\rho}{\partial t} \right] dv = 0. \tag{24d}$$

Since $\Omega$ is arbitrarily selected, the necessary condition for the above formula to be held is [14,54]

$$\left(\frac{\partial H_\rho}{\partial u_i^{(\alpha)}} + \frac{\partial L_\rho}{\partial u_i^{(\alpha)}}\right)\dot{u}_i^{(\alpha)} + \left(\frac{\partial H_\rho}{\partial \pi_i^{(\alpha)}} - \dot{u}_i^{(\alpha)}\right)\dot{\pi}_i^{(\alpha)} + \left(\frac{\partial H_\rho}{\partial u_{i,j}^{(\alpha)}} + \frac{\partial L_\rho}{\partial u_{i,j}^{(\alpha)}}\right)\dot{u}_{i,j}^{(\alpha)} + \frac{\partial H_\rho}{\partial t} + \frac{\partial L_\rho}{\partial t} = 0. \tag{25a}$$

Multiplying both sides of the above equation by $dt$, we have

$$\left(\frac{\partial H_\rho}{\partial u_i^{(\alpha)}} + \frac{\partial L_\rho}{\partial u_i^{(\alpha)}}\right)du_i^{(\alpha)} + \left(\frac{\partial H_\rho}{\partial \pi_i^{(\alpha)}} - \dot{u}_i^{(\alpha)}\right)d\pi_i^{(\alpha)} + \left(\frac{\partial H_\rho}{\partial u_{i,j}^{(\alpha)}} + \frac{\partial L_\rho}{\partial u_{i,j}^{(\alpha)}}\right)du_{i,j}^{(\alpha)} + \left(\frac{\partial H_\rho}{\partial t} + \frac{\partial L_\rho}{\partial t}\right)dt = 0. \tag{25b}$$

Since we assume that the variables $u_i^{(\alpha)}, \pi_i^{(\alpha)}, t$ and $u_{i,j}^{(\alpha)}, i,j = 1,2,3; \alpha = 1,2,...,M$, are generalized coordinate components independent of each other, and also $du_i^{(\alpha)}$, $d\pi_i^{(\alpha)}$, $dt$ and $du_{i,j}^{(\alpha)}$ are arbitrarily independent variables, followed by Arnold treatments [57], the necessary condition for Eq. (25b) to be held is

$$\begin{cases} \dot{u}_i^{(\alpha)} = \dfrac{\partial H_\rho}{\partial \pi_i^{(\alpha)}}, \\ \dfrac{\partial L_\rho}{\partial u_i^{(\alpha)}} = -\dfrac{\partial H_\rho}{\partial u_i^{(\alpha)}}, \\ \dfrac{\partial L_\rho}{\partial u_{i,j}^{(\alpha)}} = -\dfrac{\partial H_\rho}{\partial u_{i,j}^{(\alpha)}}, \\ \dfrac{\partial L_\rho}{\partial t} = -\dfrac{\partial H_\rho}{\partial t}. \end{cases} \qquad (26a)$$

For a conservative system, the total mechanical energy density is the Hamilton's density function, and

$$\frac{\partial L_\rho}{\partial t} = -\frac{\partial H_\rho}{\partial t} = 0. \qquad (26b)$$

As is assumed, the principle of energy conservation, like Hamilton's principle, may be seen as variable constraints, so that it could be used to identify independent variable couplings, which is the kernel of our proposed methodology.

Substitute Eq. (24a) into Eq. (22c), or the energy conservation equation, we obtain that

$$\iiint_\Omega \left( \frac{\partial H_\rho}{\partial u_i^{(\alpha)}} \dot{u}_i^{(\alpha)} + \frac{\partial H_\rho}{\partial \pi_i^{(\alpha)}} \dot{\pi}_i^{(\alpha)} + \frac{\partial H_\rho}{\partial u_{i,j}^{(\alpha)}} \dot{u}_{i,j}^{(\alpha)} \right) dv = 0. \qquad (27a)$$

By using the first equation in Eq. group (26a), we have

$$\iiint_\Omega \left( \frac{\partial H_\rho}{\partial u_i^{(\alpha)}} \dot{u}_i^{(\alpha)} + \dot{u}_i^{(\alpha)} \dot{\pi}_i^{(\alpha)} + \frac{\partial H_\rho}{\partial u_{i,j}^{(\alpha)}} \dot{u}_{i,j}^{(\alpha)} \right) dv = 0. \qquad (27b)$$

Using the identity equation

$$\frac{\partial H_\rho}{\partial u_{i,j}^{(\alpha)}} \dot{u}_{i,j}^{(\alpha)} = \left( \frac{\partial H_\rho}{\partial u_{i,j}^{(\alpha)}} \dot{u}_i^{(\alpha)} \right)_{,j} - \left( \frac{\partial H_\rho}{\partial u_{i,j}^{(\alpha)}} \right)_{,j} \dot{u}_i^{(\alpha)}, \qquad (27c)$$

and substituting it into Eq. (27b), we have

$$\iiint_\Omega \left[ \frac{\partial H_\rho}{\partial u_i^{(\alpha)}} \dot{u}_i^{(\alpha)} + \dot{u}_i^{(\alpha)} \dot{\pi}_i^{(\alpha)} - \left( \frac{\partial H_\rho}{\partial u_{i,j}^{(\alpha)}} \right)_{,j} \dot{u}_i^{(\alpha)} + \left( \frac{\partial H_\rho}{\partial u_{i,j}^{(\alpha)}} \dot{u}_i^{(\alpha)} \right)_{,j} \right] dv = 0. \qquad (27d)$$

In the above equation, by using Gauss's theorem to convert volume integral into enclosed surface integral, and rearranging it, we have

$$\iiint_\Omega \left[ \frac{\partial H_\rho}{\partial u_i^{(\alpha)}} \dot{u}_i^{(\alpha)} + \dot{u}_i^{(\alpha)} \dot{\pi}_i^{(\alpha)} - \left( \frac{\partial H_\rho}{\partial u_{i,j}^{(\alpha)}} \right)_{,j} \dot{u}_i^{(\alpha)} \right] dv = -\oiint_{\partial\Omega} \left( \frac{\partial H_\rho}{\partial u_{i,j}^{(\alpha)}} l_j \right) \dot{u}_i^{(\alpha)} ds. \qquad (27e)$$

Now, we prove that

$$\sigma_{ij}^{(\alpha)} = \frac{\partial H_\rho}{\partial u_{i,j}^{(\alpha)}}. \qquad (28a)$$

From Eq. (26a), we know that

$$\frac{\partial H_\rho}{\partial u_{i,j}^{(\alpha)}} = -\frac{\partial L_\rho}{\partial u_{i,j}^{(\alpha)}} = \frac{\partial V_\rho}{\partial u_{i,j}^{(\alpha)}}.$$

Furthermore, by using the relationship between stress and strain energy density function, it is easy to show that

$$\sigma_{ij}^{(\alpha)} = \frac{\partial H_\rho}{\partial u_{i,j}^{(\alpha)}}. \tag{28b}$$

By using Eq. (28a), Eq. (27e) becomes

$$\iiint_\Omega \left[ \frac{\partial H_\rho}{\partial u_i^{(\alpha)}} + \dot{\pi}_i^{(\alpha)} - \left( \frac{\partial H_\rho}{\partial u_{i,j}^{(\alpha)}} \right)_{,j} \right] \dot{u}_i^{(\alpha)} dv = - \oiint_{\partial\Omega} \left( \sigma_{ij}^{(\alpha)} l_j \right) \dot{u}_i^{(\alpha)} ds$$

As was discussed in Sec. 3.1, the same surface boundary conditions are employed, i.e., the traction-free conditions are considered to guarantee the system to be conservative, the above equation is simplified as

$$\iiint_\Omega \left[ \dot{\pi}_i^{(\alpha)} + \frac{\partial H_\rho}{\partial u_i^{(\alpha)}} - \left( \frac{\partial H_\rho}{\partial u_{i,j}^{(\alpha)}} \right)_{,j} \right] \dot{u}_i^{(\alpha)} dv = 0.$$

Arbitrary choice of integral volume $\Omega$ in the above equation implies that its integrand must vanish [14,54], namely

$$\left[ \dot{\pi}_i^{(\alpha)} + \frac{\partial H_\rho}{\partial u_i^{(\alpha)}} - \left( \frac{\partial H_\rho}{\partial u_{i,j}^{(\alpha)}} \right)_{,j} \right] \dot{u}_i^{(\alpha)} = 0.$$

Or

$$\left[ \dot{\pi}_i^{(\alpha)} + \frac{\partial H_\rho}{\partial u_i^{(\alpha)}} - \left( \frac{\partial H_\rho}{\partial u_{i,j}^{(\alpha)}} \right)_{,j} \right] du_i^{(\alpha)} = 0. \tag{29a}$$

Following the previous treatments [57], from Eq. (29a) we know that the coefficients of the independent variable $du_i^{(\alpha)}$ must vanish, i.e.,

$$\dot{\pi}_i^{(\alpha)} = -\frac{\partial H_\rho}{\partial u_i^{(\alpha)}} + \left( \frac{\partial H_\rho}{\partial u_{i,j}^{(\alpha)}} \right)_{,j}. \tag{29b}$$

Writing the first formula in Eq. group (26a) and Eq. (29b) together, we have

$$\begin{cases} \dot{u}_i^{(\alpha)} = \dfrac{\partial H_\rho}{\partial \pi_i^{(\alpha)}}, \\ \dot{\pi}_i^{(\alpha)} = -\dfrac{\partial H_\rho}{\partial u_i^{(\alpha)}} + \left( \dfrac{\partial H_\rho}{\partial u_{i,j}^{(\alpha)}} \right)_{,j}, \end{cases} \tag{30}$$

where $i, j, = 1, 2, 3$, and $\alpha = 1, 2, ..., M$, can be any number of medium compositions as expected.

Eq. group (30) is Hamilton's equations for an arbitrarily anisotropic and multiphasic porous medium in a conservative case, derived directly from the principle of conservation of energy. Exactly, it is Eq. group (13f) derived by using Hamilton's principle, both of which are validated with each other.

3.2.2 *Lagrange's equation from the principle of energy conservation for a conservative system*

Next, we formulate Lagrange's equation of an arbitrarily anisotropic and multiphasic porous medium from the perspective of energy conservation in a conservative system.

Starting from the principle of conservation of energy, considering Eq. (26b), and substituting Eq. (24c) into Eq. (22c), we have

$$\iiint_\Omega \left( \dot{u}_i^{(\alpha)} \dot{\pi}_i^{(\alpha)} - \frac{\partial L_\rho}{\partial u_i^{(\alpha)}} \dot{u}_i^{(\alpha)} - \frac{\partial L_\rho}{\partial u_{i,j}^{(\alpha)}} \dot{u}_{i,j}^{(\alpha)} \right) dv = 0. \tag{31a}$$

With the help of the following identity formula

$$\frac{\partial L_\rho}{\partial u_{i,j}^{(\alpha)}} \dot{u}_{i,j}^{(\alpha)} = \left( \frac{\partial L_\rho}{\partial u_{i,j}^{(\alpha)}} \dot{u}_i^{(\alpha)} \right)_{,j} - \left( \frac{\partial L_\rho}{\partial u_{i,j}^{(\alpha)}} \right)_{,j} \dot{u}_i^{(\alpha)}, \tag{31b}$$

Eq. (31a) is rewritten as

$$\iiint_\Omega \left[ \dot{u}_i^{(\alpha)} \dot{\pi}_i^{(\alpha)} - \frac{\partial L_\rho}{\partial u_i^{(\alpha)}} \dot{u}_i^{(\alpha)} - \left( \frac{\partial L_\rho}{\partial u_{i,j}^{(\alpha)}} \dot{u}_i^{(\alpha)} \right)_{,j} + \left( \frac{\partial L_\rho}{\partial u_{i,j}^{(\alpha)}} \right)_{,j} \dot{u}_i^{(\alpha)} \right] dv = 0.$$

By using Gauss's theorem, converting the volume integral of the third term on the left-hand side of the above equation into surface integral, yields

$$\iiint_\Omega \left[ \dot{\pi}_i^{(\alpha)} - \frac{\partial L_\rho}{\partial u_i^{(\alpha)}} + \left( \frac{\partial L_\rho}{\partial u_{i,j}^{(\alpha)}} \right)_{,j} \right] \dot{u}_i^{(\alpha)} dv = \oiint_{\partial \Omega} \left( \frac{\partial L_\rho}{\partial u_{i,j}^{(\alpha)}} l_j \right) \dot{u}_i^{(\alpha)} ds. \tag{31c}$$

Considering the third equation in Eq. group (26a) and using Eq. (28a), we have

$$\sigma_{ij}^{(\alpha)} = -\frac{\partial L_\rho}{\partial u_{i,j}^{(\alpha)}}.$$

Substituting the above equation into Eq. (31c), and using the traction free conditions as is done before for guaranteeing the system to be conservative, we have

$$\iiint_\Omega \left[ \dot{\pi}_i^{(\alpha)} - \frac{\partial L_\rho}{\partial u_i^{(\alpha)}} + \left( \frac{\partial L_\rho}{\partial u_{i,j}^{(\alpha)}} \right)_{,j} \right] \dot{u}_i^{(\alpha)} dv = 0. \tag{31d}$$

Following the previous procedure, the integrand in Eq. (31d) must vanish because of the arbitrary choice of integral volume $\Omega$ [14,54]. Therefore,

$$\left[ \dot{\pi}_i^{(\alpha)} - \frac{\partial L_\rho}{\partial u_i^{(\alpha)}} + \left( \frac{\partial L_\rho}{\partial u_{i,j}^{(\alpha)}} \right)_{,j} \right] \dot{u}_i^{(\alpha)} = 0. \tag{31e}$$

Hence,

$$\left[ \dot{\pi}_i^{(\alpha)} - \frac{\partial L_\rho}{\partial u_i^{(\alpha)}} + \left( \frac{\partial L_\rho}{\partial u_{i,j}^{(\alpha)}} \right)_{,j} \right] du_i^{(\alpha)} = 0. \tag{31f}$$

Also, following the previous treatments [57], the fact that the above equation is held with the independent variables $du_i^{(\alpha)}$ implies that their coefficients must vanish, i.e.,

$$\dot{\pi}_i^{(\alpha)} - \frac{\partial L_\rho}{\partial u_i^{(\alpha)}} + \left( \frac{\partial L_\rho}{\partial u_{i,j}^{(\alpha)}} \right)_{,j} = 0. \tag{31g}$$

Finally, substituting the definition of generalized momentum density shown in Eq. (17a) into the above equation, yields

$$\frac{d}{dt} \left( \frac{\partial L_\rho}{\partial \dot{u}_i^{(\alpha)}} \right) - \frac{\partial L_\rho}{\partial u_i^{(\alpha)}} + \left( \frac{\partial L_\rho}{\partial u_{i,j}^{(\alpha)}} \right)_{,j} = 0. \tag{32}$$

Comparing the above equation with Eq. (16), we know that Eq. (32), directly derived from the principle of energy conservation is Lagrange's equation derived from Hamilton's principle for an arbitrarily anisotropic and multiphase porous continuous conservative medium.

### 3.2.3 *The elastodynamic equation of motion from energy conservation for conservative systems*

In this section, we will formulate the dynamical equation of motion in an arbitrarily anisotropic and multiphasic porous medium, directly from the perspective of energy conservation for a conservative system.

Substituting Eqs. (2d) and (6a) for kinetic energy and potential energy density functions into the energy conservation equation in Eq. (22c) by neglecting body forces, yields

$$\frac{d}{dt}\left[\iiint_\Omega \left(\frac{1}{2}\rho_{\alpha\beta}\dot{u}_i^{(\alpha)}\dot{u}_i^{(\beta)} + \frac{1}{2}C_{ijkl}^{(\alpha,\beta)}u_{i,j}^{(\alpha)}u_{k,l}^{(\beta)}\right)dv\right] = 0. \tag{33a}$$

As is assumed, because the system is conservative, there are only kinetic energy and potential energy in the system. To guarantee Eq. (33a), we need to consider the conservative conditions, i.e., there is no energy transport across the surface of $\partial\Omega$.

As is done in Sec. 3.1, we consider the enclosed surface a traction-free surface, i.e., $\sigma_i^\alpha = \sigma_{ij}^\alpha l_j = 0$ on $\partial\Omega$. Anyway, the energy conservation equation in Eq. (33a) can be written as

$$\frac{d}{dt}\left[\iiint_\Omega \left(\frac{1}{2}\rho_{\alpha\beta}\dot{u}_i^{(\alpha)}\dot{u}_i^{(\beta)} + \frac{1}{2}C_{ijkl}^{(\alpha,\beta)}u_{i,j}^{(\alpha)}u_{k,l}^{(\beta)}\right)dv\right] = \oiint_{\partial\Omega}(\sigma_{ij}^{(\beta)}l_j\dot{u}_i^{(\beta)})ds. \tag{33b}$$

The term on the right-hand side of Eq. (33b) should be considered as the contributions to the potential energy since the stresses are related to strains with Hooke's law. It will be seen that this term is cancelled with the one on the left-hand side. The advantage of formulating dynamical equation from energy conservation does not make a big difference between the conservative and nonconservative system. As is done before, one may suppose the term on the right-hand side of Eq. (33b) vanishes so that Eq. (33b) becomes the energy conservation equation in a conservative system. In this case, the correspondent term on the left-hand side is naturally zero because of the traction free surface boundary conditions. In both cases, it turns out that they are equivalent with each other.

Starting from Eq. (33b), following the previous treatment, using Reynold's transport theorem [31], Eq. (33b) becomes

$$\iiint_\Omega (\rho_{\alpha\beta}\ddot{u}_i^{(\alpha)}\dot{u}_i^{(\beta)} + C_{ijkl}^{(\alpha,\beta)}u_{i,j}^{(\alpha)}\dot{u}_{k,l}^{(\beta)})dv = 0.$$

By using generalized Hook's law, $\sigma_{kl}^\beta = C_{ijkl}^{(\beta,\alpha)}e_{ij}^{(\alpha)}$ shown in Eq. (7), or $\sigma_{kl}^\beta = C_{ijkl}^{(\beta,\alpha)}u_{i,j}^{(\alpha)}$, the above equation is written as

$$\iiint_\Omega (\rho_{\alpha\beta}\ddot{u}_i^{(\alpha)}\dot{u}_i^{(\beta)} + \sigma_{ij}^{(\beta)}\dot{u}_{i,j}^{(\beta)})dv = 0. \tag{33c}$$

By using the identity formula

$$\sigma_{ij}^{(\beta)}\dot{u}_{i,j}^{(\beta)} = (\sigma_{ij}^{(\beta)}\dot{u}_i^{(\beta)})_{,j} - \sigma_{ij,j}^{(\beta)}\dot{u}_i^{(\beta)},$$

Eq. (33c) can be written as

$$\iiint_\Omega [\rho_{\alpha\beta}\ddot{u}_i^{(\alpha)}\dot{u}_i^{(\beta)} + (\sigma_{ij}^{(\beta)}\dot{u}_i^{(\beta)})_{,j} - \sigma_{ij,j}^{(\beta)}\dot{u}_i^{(\beta)}]dv = 0, \tag{33d}$$

or

$$\iiint_\Omega (\rho_{\alpha\beta}\ddot{u}_i^{(\alpha)} - \sigma_{ij,j}^{(\beta)})\dot{u}_i^{(\beta)}dv + \iiint_\Omega (\sigma_{ij}^{(\beta)}\dot{u}_i^{(\beta)})_{,j}dv = 0. \tag{33e}$$

According to Gauss's theorem, the volume integral of the second term on the left-hand side of the above equation is converted into the surface integral. Therefore, Eq. (33e) becomes

$$\iiint_\Omega (\rho_{\alpha\beta}\ddot{u}_i^{(\alpha)} - \sigma_{ij,j}^{(\beta)})\dot{u}_i^{(\beta)}dv + \oiint_{\partial\Omega}(\sigma_{ij}^{(\alpha)}l_j\dot{u}_i^{(\alpha)})ds = 0. \tag{33f}$$

For a conservative system, as is assumed before, the traction-free boundary conditions are considered, the above equation is simplified as

$$\iiint_\Omega (\rho_{\alpha\beta}\ddot{u}_i^{(\alpha)} - \sigma_{ij,j}^{(\beta)})\dot{u}_i^{(\beta)}dv = 0. \tag{33g}$$

As is seen, even from Eq. (33b), we still can obtain Eq. (33g) without using the traction free boundary conditions.

Following the previous procedures [14,54], to guarantee Eq. (33g) to be held for arbitrary choice of integral volume $\Omega$, its integrand must vanish, i.e.,

$$(\rho_{\alpha\beta}\ddot{u}_i^{(\alpha)} - \sigma_{ij,j}^{(\beta)})\dot{u}_i^{(\beta)} = 0. \tag{34a}$$

Also, following the previous treatments [57], from Eq. (34a), we have

$$(\rho_{\alpha\beta}\ddot{u}_i^{(\alpha)} - \sigma_{ij,j}^{(\beta)})du_i^{(\beta)} = 0. \tag{34b}$$

Hence,

$$\rho_{\alpha\beta}\ddot{u}_i^{(\alpha)} - \sigma_{ij,j}^{(\beta)} = 0. \tag{34c}$$

Exchanging $\alpha$ and $\beta$ in the above formula, and considering the symmetry of $\rho_{\alpha\beta} = \rho_{\beta\alpha}$, we have

$$\rho_{\alpha\beta}\ddot{u}_i^{(\beta)} - \sigma_{ij,j}^{(\alpha)} = 0, \alpha, \beta = 1, 2, ..., M; i, j = 1, 2, 3, \tag{34d}$$

where $M$ represents the total number of material compositions in the porous medium of interest.

By examining Eq. (34d), we find that it is exactly the elastodynamic equation of motion in an arbitrarily anisotropic and multiphasic porous medium given by Lagrange's equation in Eq. (20a) which is derived from Hamilton's principle in Sec. 3.1.

### 3.2.4 *General equations from the principle of energy conservation for nonconservative systems*

If there are non-conservative physical and dissipative forces in the medium, we can also establish the equation of motion in the same way as before. For non-conservative systems, Hamilton's principle cannot be used directly, while the proposed methodology based on the principle of energy conservation is not subject to this restriction. That is, the principle of energy conservation can also be used for formulating general mechanical equations even if

$$\frac{\partial H_\rho}{\partial t} \neq 0 \text{ and } \frac{\partial L_\rho}{\partial t} \neq 0.$$

If there exists this situation, usually

$$\frac{\partial x_i}{\partial t} \neq 0,$$

and generalized coordinate transformation may be introduced to get rid of constraints of displacements as is done in our previous work [44]. As is known [56],

$$\frac{du_i(x_j,t)}{dt} = \frac{\partial u_i}{\partial x_j}\frac{\partial x_j}{\partial t} + \frac{\partial u_i}{\partial t}.$$

In our discussions, as is assumed, only the linear case is considered. Therefore, high-order contributions are neglected such that

$$\frac{\partial u_i^{(\alpha)}}{\partial x_j}\frac{\partial x_j}{\partial t}=0.$$

In our discussions, since space coordinate variables and time are independently for solid media, we consider

$$\frac{\partial x_j}{\partial t}=0,$$

so that

$$\frac{du_i^{(\alpha)}(x_j,t)}{dt}=\frac{\partial u_i^{(\alpha)}(x_j,t)}{\partial t},$$

which is assumed in Sec. 2. Therefore, we have

$$\frac{\partial H_\rho}{\partial t}=0, \text{ and } \frac{\partial L_\rho}{\partial t}=0.$$

In this case, the energy conservation equation in Eq. (21) can be written as

$$\frac{d}{dt}\left(\iiint_\Omega E_\rho dv\right)=\oiint_{\partial\Omega}(\sigma_{ij}^{(\alpha)}l_j\dot{u}_i^{(\alpha)})ds+\iiint_\Omega f_i^{(\alpha)}\dot{u}_i^{(\alpha)}dv+\iiint_\Omega F_i^{(\alpha)}\dot{u}_i^{(\alpha)}dv, \tag{35a}$$

where $f_i^{(\alpha)}$ and $F_i^{(\alpha)}$ are the equivalent body and dissipation forces in Composition $\alpha$ in the porous media, respectively. The first, second, and third terms on the right-hand side of the above equation are the work done by the surface forces, body forces, and dissipation forces per unit time, respectively.

On the one hand, for arbitrarily anisotropic and multiphasic porous media, the dissipative energy density function can be written as [48]

$$\phi_d=\frac{1}{2}b_{\alpha\beta}\dot{w}_i^{(\alpha,\beta)}\dot{w}_i^{(\alpha,\beta)}, i=1,2,3;\alpha,\beta=1,2,...,M. \tag{35b}$$

where $\dot{w}_i^{(\alpha,\beta)}=\dot{u}_i^{(\alpha)}-\dot{u}_i^{(\beta)}$ are the relative velocity components between Composition $\alpha$ and Composition $\beta$ in the $i-$ direction; $b_{\alpha\beta}$ are the components of the symmetric friction matrix $B$ [48], and $b_{\alpha\beta}=b_{\beta\alpha}$. The friction coefficient represents the friction force generated by the relative motion when the velocity vectors of the two media are inconsistent, and it is assumed that the friction coefficients of the relative velocities in the three directions are consistent. If they are inconsistent, the friction coefficient is anisotropic, so the friction coefficients should be treated differently in different directions. Also, the friction coefficients are related to the direction of the relative velocity, so it should be written as $b_{\alpha\beta}^{(i,j)}$. In our work, only the isotropic friction coefficients are concerned and $b_{\alpha\beta}^{(i,j)}$ is denoted simply with $b_{\alpha\beta}^{(\gamma)}$ as $b_{\alpha\beta}^{(i,j)}=b_{\alpha\beta}^{(j,i)}$. From Eq. (35b), it is seen that the dissipation forces can be written as [48]

$$F_i^{(\alpha)}=-b_{\gamma\beta}^\alpha \dot{w}_i^{(\gamma,\beta)}, i=1,2,3;\alpha,\beta,\gamma=1,2,...,M. \tag{35c}$$

However, considering the isotropic properties of the friction coefficients does not mean that the treatment loses its generality since it only influences the representation of $F_i^{(\alpha)}$. From the energy conservation equation shown in Eq. (35a), we have

$$\iiint_\Omega\left(\frac{dT_\rho}{dt}+\frac{dV_\rho}{dt}\right)dv=\oiint_{\partial\Omega}(\sigma_{ij}^{(\alpha)}l_j\dot{u}_i^{(\alpha)})ds+\iiint_\Omega f_i^{(\alpha)}\dot{u}_i^{(\alpha)}dv+\iiint_\Omega F_i^{(\alpha)}\dot{u}_i^{(\alpha)}dv, \tag{35d}$$

The above equation can be used to derive Lagrange's equation and the elastodynamic equation of motion in a continuous nonconservative system.

On the one hand, the rate of kinetic energy density with time is written as

$$\frac{dT_\rho}{dt} = \frac{\partial T_\rho}{\partial u_i^{(\alpha)}} \dot{u}_i^{(\alpha)} + \frac{\partial T_\rho}{\partial \dot{u}_i^{(\alpha)}} \ddot{u}_i^{(\alpha)} + \frac{\partial T_\rho}{\partial u_{i,j}^{(\alpha)}} \dot{u}_{i,j}^{(\alpha)}. \tag{35e}$$

By using the identity formula

$$\frac{\partial T_\rho}{\partial \dot{u}_i^{(\alpha)}} \ddot{u}_i^{(\alpha)} = \frac{d}{dt}\left(\frac{\partial T_\rho}{\partial \dot{u}_i^{(\alpha)}} \dot{u}_i^{(\alpha)}\right) - \frac{d}{dt}\left(\frac{\partial T_\rho}{\partial \dot{u}_i^{(\alpha)}}\right)\dot{u}_i^{(\alpha)},$$

Eq. (35e) can be written as

$$\frac{dT_\rho}{dt} = \frac{\partial T_\rho}{\partial u_i^{(\alpha)}} \dot{u}_i^{(\alpha)} + \frac{d}{dt}\left(\frac{\partial T_\rho}{\partial \dot{u}_i^{(\alpha)}} \dot{u}_i^{(\alpha)}\right) - \frac{d}{dt}\left(\frac{\partial T_\rho}{\partial \dot{u}_i^{(\alpha)}}\right)\dot{u}_i^{(\alpha)} + \frac{\partial T_\rho}{\partial u_{i,j}^{(\alpha)}} \dot{u}_{i,j}^{(\alpha)}.$$

According to the definition of kinetic energy density, the above equation can be simplified as

$$\frac{dT_\rho}{dt} = \frac{\partial T_\rho}{\partial u_i^{(\alpha)}} \dot{u}_i^{(\alpha)} + 2\frac{dT_\rho}{dt} - \frac{d}{dt}\left(\frac{\partial T_\rho}{\partial \dot{u}_i^{(\alpha)}}\right)\dot{u}_i^{(\alpha)} + \frac{\partial T_\rho}{\partial u_{i,j}^{(\alpha)}} \dot{u}_{i,j}^{(\alpha)}, \tag{35f}$$

where the identity

$$\frac{d}{dt}\left(\frac{\partial T_\rho}{\partial \dot{u}_i^{(\alpha)}} \dot{u}_i^{(\alpha)}\right) = \frac{2dT_\rho}{dt},$$

has been used. Therefore, Eq. (35f) can be written as

$$\frac{dT_\rho}{dt} = \frac{d}{dt}\left(\frac{\partial T_\rho}{\partial \dot{u}_i^{(\alpha)}}\right)\dot{u}_i^{(\alpha)} - \frac{\partial T_\rho}{\partial u_i^{(\alpha)}} \dot{u}_i^{(\alpha)} - \frac{\partial T_\rho}{\partial u_{i,j}^{(\alpha)}} \dot{u}_{i,j}^{(\alpha)} \tag{35g}$$

Since the elastic potential energy density is not a function of particle velocity and time explicitly, the above equation can be written as

$$\frac{dT_\rho}{dt} = \frac{d}{dt}\left(\frac{\partial L_\rho}{\partial \dot{u}_i^{(\alpha)}}\right)\dot{u}_i^{(\alpha)} - \frac{\partial T_\rho}{\partial u_i^{(\alpha)}} \dot{u}_i^{(\alpha)} - \frac{\partial T_\rho}{\partial u_{i,j}^{(\alpha)}} \dot{u}_{i,j}^{(\alpha)} \tag{35h}$$

On the other hand, the rate of elastic potential energy density with time can be written as

$$\frac{dV_\rho}{dt} = \frac{\partial V_\rho}{\partial u_i^{(\alpha)}} \dot{u}_i^{(\alpha)} + \frac{\partial V_\rho}{\partial u_{i,j}^{(\alpha)}} \dot{u}_{i,j}^{(\alpha)}. \tag{35i}$$

From Eqs. (35h) and (35i), we have

$$\frac{dT_\rho}{dt} + \frac{dV_\rho}{dt} = \frac{d}{dt}\left(\frac{\partial L_\rho}{\partial \dot{u}_i^{(\alpha)}}\right)\dot{u}_i^{(\alpha)} - \frac{\partial L_\rho}{\partial u_i^{(\alpha)}} \dot{u}_i^{(\alpha)} - \frac{\partial L_\rho}{\partial u_{i,j}^{(\alpha)}} \dot{u}_{i,j}^{(\alpha)}. \tag{36a}$$

By using Eq. (10a), for an arbitrarily anisotropic and multiphasic porous medium, the Hamilton's density function can be written as

$$H_\rho(u_i^{(\alpha)}, \dot{u}_i^{(\alpha)}, u_{i,j}^{(\alpha)}) = \pi_i^{(\alpha)} \dot{u}_i^{(\alpha)} - L_\rho(u_i^{(\alpha)}, \pi_i^{(\alpha)}, u_{i,j}^{(\alpha)}), i,j = 1,2,3; \alpha = 1,2,...,M. \tag{36b}$$

Following Arnold's treatments [57], the time derivative of Hamilton's density function is written as

$$dH_\rho = \frac{\partial H_\rho}{\partial u_i^{(\alpha)}} du_i^{(\alpha)} + \frac{\partial H_\rho}{\partial \pi_i^{(\alpha)}} d\pi_i^{(\alpha)} + \frac{\partial H_\rho}{\partial u_{i,j}^{(\alpha)}} du_{i,j}^{(\alpha)} + \frac{\partial H_\rho}{\partial t} dt. \tag{36c}$$

Using Legendre transformation, we also have,

$$dH_\rho = d(\pi_i^{(\alpha)}\dot{u}_i^{(\alpha)} - L_\rho) = \dot{u}_i^{(\alpha)}d\pi_i^{(\alpha)} + \pi_i^{(\alpha)}d\dot{u}_i^{(\alpha)} - \frac{\partial L_\rho}{\partial u_i^{(\alpha)}}du_i^{(\alpha)} - \frac{\partial L_\rho}{\partial \dot{u}_i^{(\alpha)}}d\dot{u}_i^{(\alpha)} - \frac{\partial L_\rho}{\partial u_{i,j}^{(\alpha)}}du_{i,j}^{(\alpha)} - \frac{\partial L_\rho}{\partial t}dt.$$

By using Eq. (17a) for the definition of generalized momentum, the above equation is simplified into

$$dH_\rho = \dot{u}_i^{(\alpha)}d\pi_i^{(\alpha)} - \frac{\partial L_\rho}{\partial u_i^{(\alpha)}}du_i^{(\alpha)} - \frac{\partial L_\rho}{\partial u_{i,j}^{(\alpha)}}du_{i,j}^{(\alpha)} - \frac{\partial L_\rho}{\partial t}dt. \tag{36d}$$

Comparing Eq. (36c) with Eq. (36d), yields

$$\begin{cases} \dot{u}_i^{(\alpha)} = \dfrac{\partial H_\rho}{\partial \pi_i^{(\alpha)}}, \\[4pt] \dfrac{\partial L_\rho}{\partial u_i^{(\alpha)}} = -\dfrac{\partial H_\rho}{\partial u_i^{(\alpha)}}, \\[4pt] \dfrac{\partial L_\rho}{\partial u_{i,j}^{(\alpha)}} = -\dfrac{\partial H_\rho}{\partial u_{i,j}^{(\alpha)}}, \\[4pt] \dfrac{\partial L_\rho}{\partial t} = -\dfrac{\partial H_\rho}{\partial t}. \end{cases} \tag{36e}$$

Substituting the second and third equations in Eq. group (36e) into Eq. (36a), yields

$$\frac{d(T_\rho + V_\rho)}{dt} = \dot{\pi}_i^{(\alpha)}\dot{u}_i^{(\alpha)} + \frac{\partial H_\rho}{\partial u_i^{(\alpha)}}\dot{u}_i^{(\alpha)} + \frac{\partial H_\rho}{\partial u_{i,j}^{(\alpha)}}\dot{u}_{i,j}^{(\alpha)}. \tag{37a}$$

Therefore, the energy conservation equation in Eq. (35d) can be rewritten as

$$\iiint_\Omega \left( \dot{\pi}_i^{(\alpha)}\dot{u}_i^{(\alpha)} + \frac{\partial H_\rho}{\partial u_i^{(\alpha)}}\dot{u}_i^{(\alpha)} + \frac{\partial H_\rho}{\partial u_{i,j}^{(\alpha)}}\dot{u}_{i,j}^{(\alpha)} \right) dv = \oiint_{\partial\Omega} (\sigma_{ij}^{(\alpha)}l_j\dot{u}_i^{(\alpha)})ds + \iiint_\Omega f_i^{(\alpha)}\dot{u}_i^{(\alpha)}dv + \iiint_\Omega F_i^{(\alpha)}\dot{u}_i^{(\alpha)}dv. \tag{37b}$$

Now, substituting the identity equation in Eq. (27c) into the above equation and converting the part of the volume integral into the surface one by using the Gauss's theorem, i.e.,

$$\iiint_\Omega \left( \frac{\partial H_\rho}{\partial u_{i,j}^{(\alpha)}}\dot{u}_i^{(\alpha)} \right)_{,j} dv = \oiint_{\partial\Omega} \left( \frac{\partial H_\rho}{\partial u_{i,j}^{(\alpha)}}l_j \right)\dot{u}_i^{(\alpha)}ds,$$

Eq. (37b) can be simplified as

$$\iiint_\Omega \left[ \left( \dot{\pi}_i + \frac{\partial H_\rho}{\partial u_i^{(\alpha)}} \right)\dot{u}_i^{(\alpha)} - \left( \frac{\partial H_\rho}{\partial u_{i,j}^{(\alpha)}} \right)_{,j} \dot{u}_i^{(\alpha)} \right] dv = \oiint_{\partial\Omega} \left( \sigma_{ij}^{(\alpha)} - \frac{\partial H_\rho}{\partial u_{i,j}^{(\alpha)}} \right) l_j \dot{u}_i^{(\alpha)} ds + \iiint_\Omega \left( f_i^{(\alpha)} + F_i^{(\alpha)} \right)\dot{u}_i^{(\alpha)} dv. \tag{37c}$$

Using Eq. (28a), the above equation can be further simplified as

$$\iiint_\Omega \left[ \left( \dot{\pi}_i + \frac{\partial H_\rho}{\partial u_i^{(\alpha)}} \right) - \left( \frac{\partial H_\rho}{\partial u_{i,j}^{(\alpha)}} \right)_{,j} - f_i^{(\alpha)} - F_i^{(\alpha)} \right] \dot{u}_i^{(\alpha)} dv = 0. \tag{37d}$$

Following the previous procedures [14,54], it is easy to show that one Hamilton's equation in an arbitrarily anisotropic and multiphase porous medium is formulated as

$$\dot{\pi}_i = -\frac{\partial H_\rho}{\partial u_i^{(\alpha)}} + \left( \frac{\partial H_\rho}{\partial u_{i,j}^{(\alpha)}} \right)_{,j} + f_i^{(\alpha)} + F_i^{(\alpha)}. \tag{37e}$$

With Eq. group (36e) and Eq. (37e), Hamilton's equations in nonconservative system are formulated, and are rewritten as

$$\begin{cases} \dot{u}_i^{(\alpha)} = \dfrac{\partial H_\rho}{\partial \pi_i^{(\alpha)}}, \\ \dot{\pi}_i = -\dfrac{\partial H_\rho}{\partial u_i^{(\alpha)}} + \left(\dfrac{\partial H_\rho}{\partial u_{i,j}^{(\alpha)}}\right)_{,j} + f_i^{(\alpha)} + F_i^{(\alpha)}. \end{cases} \tag{37f}$$

By using Eq. (37e), Eqs. group (36e), and the definition of generalized momentum density in Eq. (10b), the Lagrange's equation is obtained in the form of

$$\dfrac{d}{dt}\left(\dfrac{\partial L_\rho}{\partial \dot{u}_i^{(\alpha)}}\right) - \dfrac{\partial L_\rho}{\partial u_i^{(\alpha)}} + \left(\dfrac{\partial L_\rho}{\partial u_{i,j}^{(\alpha)}}\right)_{,j} - f_i^{(\alpha)} - F_i^{(\alpha)} = 0. \tag{37g}$$

Since

$$\dfrac{\partial L_\rho}{\partial \dot{u}_i^{(\alpha)}} = \dfrac{\partial}{\partial \dot{u}_i^{(\alpha)}}\left(\dfrac{1}{2}\rho_{\alpha\beta}\dot{u}_i^{(\alpha)}\dot{u}_i^{(\beta)}\right) = \rho_{\alpha\beta}\dot{u}_i^{(\beta)}, \quad \sigma_{ij}^{(\alpha)} = -\dfrac{\partial L_\rho}{\partial u_{i,j}^{(\alpha)}}, \quad \text{and} \quad \dfrac{\partial L_\rho}{\partial u_i^{(\alpha)}} = 0,$$

Substituting the above three equations into Eq. (37g) yield

$$\rho_{\alpha\beta}\ddot{u}_i^{(\beta)} - \sigma_{ij,j}^{(\alpha)} - f_i^{(\alpha)} - F_i^{(\alpha)} = 0, \tag{38a}$$

which is the general dynamical equation of motion in the arbitrarily anisotropic and multiphasic porous media. If the body forces and dissipative forces are neglected, Eq. (38a) will degenerate into Eq. (34c) for the conservative case.

Substituting Eq.(35c) into Eq.(38a), yields

$$\rho_{\alpha\beta}\ddot{u}_i^{(\beta)} - \sigma_{ij,j}^{(\alpha)} - f_i^{(\alpha)} + b_{\gamma\beta}^\alpha \dot{w}_i^{(\gamma,\beta)} = 0. \tag{38b}$$

Expanding and rearranging Eq. (38b) for a specific multiphase porous model with three compositions, we have

$$\begin{cases} \rho_{11}\ddot{u}_i^{(1)} + \rho_{12}\ddot{u}_i^{(2)} + \rho_{13}\ddot{u}_i^{(3)} - \sigma_{i,jj}^{(1)} - f_i^{(1)} + b_{12}(u_i^{(1)} - u_i^{(2)}) + b_{13}(u_i^{(1)} - u_i^{(3)}) = 0, \\ \rho_{21}\ddot{u}_i^{(1)} + \rho_{22}\ddot{u}_i^{(2)} + \rho_{23}\ddot{u}_i^{(3)} - \sigma_{i,jj}^{(2)} - f_i^{(2)} + b_{12}(u_i^{(2)} - u_i^{(1)}) + b_{23}(u_i^{(2)} - u_i^{(3)}) = 0, \\ \rho_{31}\ddot{u}_i^{(1)} + \rho_{32}\ddot{u}_i^{(2)} + \rho_{33}\ddot{u}_i^{(3)} - \sigma_{i,jj}^{(3)} - f_i^{(3)} + b_{13}(u_i^{(3)} - u_i^{(1)}) + b_{23}(u_i^{(3)} - u_i^{(2)}) = 0. \end{cases} \tag{39}$$

In the above equations, since $b_{mn}^{(\alpha)} = b_{nm}^{(\alpha)}$, which is an isotropic case for dissipation in the porous medium, its value is only identified by Medium compositions $m$ and $n$. Therefore, its superscript $\alpha$ can be omitted. If the body forces are not considered, the above equation degenerates to the dynamical equations of motion which are consistent with those in Leclaire [31] and Liu, Zhang et al [58].

Eq. (37g) is Lagrange's equation formulated from the energy conservation equation for an arbitrarily anisotropic and multiphasic porous medium, while the general elastodynamic equation of motion for the medium is shown in Equation (38a). If the external body forces and dissipation forces are not considered, Eq. (37g) will degenerate into Lagrange's equation seen in Eq. (32), which is the same as the one given by Hamilton's principles in Eq. (16) in a conservative system, while Eq. (38a) will be consistent with Eq.(20a) also given by Hamilton's principle.

Until now, Hamilton's equations, Lagrange's equations, and elastodynamic equations of motion in an arbitrarily anisotropic and multiphasic porous medium have been formulated by only using the principle of energy conservation，while the derived Lagrange's equation from energy conservation equation implies Hamilton's principle. The elastodynamic equations are also formulated directly from the axiom of energy conservation. Hamilton's principle and the principle of energy conservation in a nonconservative continuous system are compared in the followings:

No.1. Hamilton's principle in a nonconservative system is written as [6]

$$\int_{t_1}^{t_2} dt [\iiint_\Omega \delta(T_\rho - V_\rho) + \delta W_e] dv = 0,$$

with the time terminal constrains of $\delta u_i = 0$ at $t_1$ and $t_2$ respectively, where the work done for the system by external forces is denoted as $W_e$.

No.2. The principle of energy conservation in a nonconservative system is written as

$$\frac{d}{dt}\left[\iiint_\Omega (T_\rho + V_\rho - W_e) dv\right] = 0.$$

It is apparent that the proposed methodology is much simpler with clear physical meanings and without time terminal constraints. All the other assumptions are the same. From our discussions, it is shown that Hamilton's principle is equivalent with the principle of energy conservation, which implies that the physical essence of Hamilton's principle is energy conservation.

## 4. Boundary conditions at an interface from the principle of energy conservation

In this section, we will show, as an example of the application of our proposed methodology, how it can be applied to the boundary continuity conditions in two multiphasic porous media.

The kinematic and dynamical boundary conditions of multiphase porous media are important for correctly describing the reflection and transmission of acoustic waves on such boundaries. Previously, the boundary conditions of elastic media are determined by the physical states of stress and displacement. For a multiphase porous model, it is much more complicated to directly use the concepts of force and displacements to determine the conditions at an interface because the compositions and contents of the two media divided by the interface may be different. The interface conditions of Deresiewicz and Skalak [76] have been now widely used in wave propagation studies. However, some issues related to the interface conditions in porous media are still under discussion [77], and more comprehensive reviews on the boundary conditions for porous media can be seen in Carcione's work [48].

As is done before for dynamical equation formulations, the principle of energy conservation can also be used to determine the boundary conditions at an interface [76], and it can be extended to an arbitrarily anisotropic and multiphasic porous medium interface.

For a given volume domain $\Omega$ in a multiphase porous medium, regardless of the physical force, the rate of the total energy with time in the volume can be written as

$$I = \frac{d}{dt}\left[\iiint_\Omega (T_\rho + V_\rho + \phi_\rho) dv\right] = \frac{d}{dt}\iiint_\Omega \left[\frac{\rho_{\alpha\beta}}{2}\dot{u}_i^{(\alpha)}\dot{u}_i^{(\beta)} + \frac{C_{ijkl}^{(\alpha,\beta)}}{2}e_{ij}^{(\alpha)}e_{kl}^{(\beta)} - \frac{b_{\alpha\beta}}{2}(\dot{w}_i^{(\alpha,\beta)})^2\right] dv,$$

where the dissipation energy is concerned, and the friction coefficients are isotropic. Since we are only interested in boundary conditions, it does not make different if the friction coefficients are anisotropic or not. According to the principle of energy conservation, the rate of the energy with time in the volume domain is equal to the rate of work with time done by the surface force on the area of the volume enclosed. Its differential form in the two-phase and three-component porous medium can be obtained with Eq. (38b) without body forces, i.e.,

$$\rho_{\alpha\beta}\ddot{u}_i^{(\beta)} - \sigma_{ij,j}^{(\alpha)} + b_{\gamma\beta}^{(\alpha)}\dot{w}_i^{(\gamma,\beta)} = 0, i,j,\alpha,\beta,\gamma = 1,2,3. \tag{40a}$$

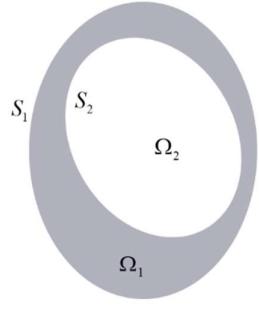

**Fig. 3.** Diagram for deriving boundary conditions from energy conservation.

Now, we consider two volumes $\Omega_1$ and $\Omega_2$ as shown in Fig. 3. $\Omega_1$ and $\Omega_2$ share with the same surface $S_2$, while $\Omega_2$ is enclosed by $S_2$, like a solid ball with its surface; while $\Omega_1$ is the volume between the outer closed surface $S_1$ and the inner closed surface $S_2$. We consider the energy conservation equations in $\Omega_1$ and $\Omega_2$ uniformly expressed as

$$\iiint_{\Omega_k} \frac{d}{dt}\left[\frac{\rho^{(\eta)}_{\alpha\beta}}{2}\dot{u}_i^{(\beta,\eta)}\dot{u}_i^{(\beta,\eta)} + \frac{C^{(\alpha,\beta)}_{ijkl}}{2}u_{ij}^{(\alpha,\eta)}u_{kl}^{(\beta,\eta)} + \frac{b_{\alpha\beta}}{2}(\dot{u}_i^{(\alpha,\eta)} - \dot{u}_i^{(\beta,\eta)})^2\right]dv = (-1)^{\eta+1}\oiint_{S_{(k)}}(\sigma_{ij}^{(\beta,\eta)}l_j^{(k)}\dot{u}_i^{(\beta,\eta)})ds, \quad (40b)$$

where $\eta = 1, 2$ are respectively expressed for the case of $\Omega_1$ and $\Omega_2$, while $S_{(1)} = S_1 + S_2$, $S_{(2)} = S_2$. It should be noted that $\eta$ index does not conform to the Einstein summation convention. Since we consider the boundary problem on the surface $S_2$, it may be set $S_1$ as free boundary conditions, and $\Omega_1$ and $\Omega_2$ don't contain any other non-conservative physical forces except the dissipation force.

According to the conditions mentioned above, due to the conservation of energy, the rate of transformation of the total energy in $\Omega_1$ and $\Omega_2$ over time is equal, i.e.

$$\oiint_{S_{(1)}}(\sigma_{ij}^{(\beta,1)}l_j\dot{u}_i^{(\beta,1)})ds = -\oiint_{S_2}(\sigma_{ij}^{(\beta,2)}l_j^{(2)}\dot{u}_i^{(\beta,2)})ds.$$

Considering the traction-free on the surface $S_1$, the above equation can be written as

$$\oiint_{S_2}(\sigma_{ij}^{(\beta,1)}l_j^{(1)}\dot{u}_i^{(\beta,1)})ds + \oiint_{S_2}(\sigma_{ij}^{(\beta,2)}l_j^{(2)}\dot{u}_i^{(\beta,2)})ds = 0. \quad (40c)$$

For any point on the surface, we have $l_j^{(1)} = -l_j^{(2)}$. Therefore, Eq. (40c) can be written as

$$\oiint_{S_2}(\sigma_{ij}^{(\beta,1)}\dot{u}_i^{(\beta,1)} - \sigma_{ij}^{(\beta,2)}\dot{u}_i^{(\beta,2)})l_j^{(1)}ds = 0. \quad (41a)$$

Since $S_2$ is arbitrarily selected in $\Omega_1$, it can be obtained from the above equation that

$$\sigma_{ij}^{(\beta,1)}l_j^{(1)}\dot{u}_i^{(\beta,1)} - \sigma_{ij}^{(\beta,2)}l_j^{(1)}\dot{u}_i^{(\beta,2)} = 0. \quad (41b)$$

Therefore, the general stress and velocity continuity boundary conditions for two different multiphase porous media are determined by the energy conservation equation. The boundary conditions for special cases are discussed in detail in the following sections.

### 4.1 Boundary conditions for solid media

Consider that the two solid media have no relative motion on the boundary, so the displacement is continuous. In this case, Eq. (41b) can be written as

$$\sigma_{ij}^{(1,1)}l_j^{(1)}\dot{u}_i^{(1,1)} - \sigma_{ij}^{(2,2)}l_j^{(1)}\dot{u}_i^{(2,2)} = 0,$$

and by using the condition $\dot{u}_i^{(1,1)} = \dot{u}_i^{(2,2)} = \dot{u}_i$ on the boundary, noting that $l_j^{(1)} = l_j$, we have

$$(\sigma_{ij}^{(1,1)} l_j - \sigma_{ij}^{(2,2)} l_j)\dot{u}_i = 0. \tag{42a}$$

Following the previous procedures, the above equation is written as

$$(\sigma_{ij}^{(1,1)} l_j - \sigma_{ij}^{(2,2)} l_j) du_i = 0. \tag{42b}$$

Hence, $\sigma_{ij}^{(1,1)} l_j = \sigma_{ij}^{(2,2)} l_j$. Let $T_i^{(1)} = \sigma_{ij}^{(1,1)} l_j$, $T_i^{(2)} = \sigma_{ij}^{(2,2)} l_j$. Eventually, the boundary conditions for stress and displacement at the interface are identified as

$$\begin{cases} T_i^{(1)} = T_i^{(2)}, \\ u_i^{(1)} = u_i^{(2)}, \end{cases} \tag{42c}$$

which are the typical stress and displacement boundary conditions [8].

### 4.2 Boundary conditions for fluid and solid media

Now, we consider the situation on the fluid-solid boundary. In this case, Eq. (42b) will be satisfied if the following equations are held.

$$\begin{cases} T_i^{(1)} = T_i^{(2)}, \\ u_i^{(1)} = u_i^{(2)}, \end{cases} \tag{42d}$$

where $T_i^{(\gamma)} = \sigma_{ij}^{(\beta,\gamma)} l_j^{(1)}$, and $u_i^{(\gamma)} = u_i^{(\beta,\gamma)} \gamma = 1, 2$, are the stress and displacement components in the outer normal direction $l_j^{(1)}$ on $S_2$, and $\beta$ is omitted because there is only one medium composition on each side of $S_2$.

Now let's discuss the conditions for satisfying Eq. (42d), which are the same as those in Eq. (42c). Since it is a fluid-solid interface, the displacement of the fluid and the solid is not necessarily continuous as it can no longer be assumed that the fluid-solid moves together in all directions on the interface. Therefore, the solid-solid boundary condition determined according to Eq. (42c) or (42d) is no longer applicable.

As we know, on the solid-solid boundary since the stress and displacement components are respectively continuous, the mass of the medium is naturally conserved, while on the fluid-solid interface, the displacement components are not necessarily equal one-to-one, otherwise, the mass is not conserved. Therefore, mass conservation can be used to give the constraint relations between displacements on the boundary [76].

When it comes to fluids (especially gases), the density of which is time dependent. However, mass conservation requires that the density within the volume element changes over time and that any unbalanced mass flows into and out of the balance of the accumulation of mass inside the volume element. The mass conservation can be expressed as [55,56]

$$\iiint_\Omega \left(\frac{\partial \rho}{\partial t}\right) dv = -\oiint_{\partial\Omega} \rho \dot{u}_i l_i ds. \tag{42e}$$

Using Gauss's theorem, the above equation can be simplified to

$$\iiint_\Omega \left[\frac{\partial \rho}{\partial t} + (\rho \dot{u}_i)_{,i}\right] dv = 0.$$

We may obtain the differential equation of mass conservation in a continuum from the arbitrary choice of the integral volume domain in the above formula [32]

$$\frac{\partial \rho}{\partial t} + \nabla \cdot (\rho \dot{\vec{u}}) = 0. \tag{42f}$$

From the above equations, we conclude that the density of both fluid and solid is independent of time, so $\partial \rho / \partial t = 0$. Therefore, the mass continuity boundary conditions can be written as

$$\nabla \cdot (\rho \dot{u}) = 0.$$

Since the particle vibration of the medium has the same velocity on both sides of the interface in the tangential direction, it does not need to be considered. On the other hand, since the densities of the two media are constant, it can be seen from the above formula that in the direction normal to the interface, the particle velocities of the two media have the difference between the projections of the vibration velocities on the interface and normal to it, and the difference should be zero, i.e.,

$$(\dot{u}_i^{(1)} - \dot{u}_i^{(2)})l_i = 0, \tag{42g}$$

which is expressed as

$$\dot{u}_n^{(1)} - \dot{u}_n^{(2)} = 0. \tag{42h}$$

where $\dot{u}_n^{(1)}$ and $\dot{u}_n^{(2)}$ are respectively the components of the particle vibration velocity of the two media normal to the interface. We may change them into displacements, so the particle displacements should have a constraint

$$u_n^{(1)} - u_n^{(2)} = 0. \tag{42i}$$

Substitute Eq. (40h) into Eq. (40d), and we have

$$[\sigma_{ij}^{(1)} - \delta_{ij}\sigma^{(2)}]l_j = 0. \tag{42j}$$

Therefore, the stress at the fluid-solid interface is $T_i^{(1)} = \delta_{ij}\sigma^{(2)}l_j$, where $i = 1$ is set to the normal direction of the interface, and $\sigma^{(2)}$ is the stress in the fluid, or

$$\begin{cases} T_1^{(1)} = \sigma^{(2)} = -P_f, \\ T_2^{(1)} = 0, \\ T_3^{(1)} = 0, \end{cases} \tag{42k}$$

where $T_1^{(1)}$ is assumed to be the normal component of stress at the interface, $P_f$ is the fluid pressure, $T_2^{(1)}$ and $T_3^{(1)}$ are the two tangential stress components on the interface. Eqs. (42i) and (42k) are our commonly used solid-fluid boundary conditions for stress and displacement at the interface [8]. These boundary conditions satisfy both energy and mass conservations.

### 4.3 Boundary conditions for porous media

If there are two different porous media saturated with a single fluid, it is known from Eq. (41b) that,

$$T_i^{(\beta,1)}\dot{u}_i^{(\beta,1)} = T_i^{(\beta,2)}\dot{u}_i^{(\beta,2)}. \tag{43a}$$

In the above formula $T_i^{(\beta,\gamma)} = \sigma_{ij}^{(\beta,\gamma)}l_j^{(1)}, \gamma = 1,2$, are the stress components in the outward normal direction $l_j^{(1)}$ on $S_2$. Now let's discuss the condition for satisfying Eq. (41a).

The first condition is that Eq. (43a) will be satisfied if $T_i^{(\beta,1)} = T_i^{(\beta,2)}$, and $\dot{u}_i^{(\beta,1)} = \dot{u}_i^{(\beta,2)}$. These conditions imply that there is no relative motion between the pore fluid and the solid, which is not true. Therefore, the constraint of the fluid velocity is incorrect [33]. Due to the relative motion of the velocity between the fluid-solid medium, the motion of the constraint condition on the boundary should satisfy the conservation of mass as is just discussed for the case of fluid-solid boundary conditions. In this case, Eq. (43a) can be written as

$$(\sigma_{ij}^{(1)} + \delta_{ij}\sigma^{(1)})l_j\dot{u}_i^{(1)} - \sigma^{(1)}\dot{w}_i^{(1)}l_i = (\sigma_{ij}^{(2)} + \delta_{ij}\sigma^{(2)})l_j\dot{u}_i^{(2)} - \sigma^{(2)}\dot{w}_i^{(2)}l_i.$$

In the above equation, the superscript $\beta$ is omitted, and $\gamma = 1,2$, $\sigma_{ij}^{(\gamma)}$ and $\sigma^{(\gamma)}$ represent the stress components in the skeleton and the fluid of the porous media, respectively, and $\dot{u}_i^{(\gamma)}$ and $\dot{w}_i^{(\gamma)}$ represent the solid skeleton velocity and the velocity of the solid relative to the pore fluid, respectively. Following the analyses of the case for a fluid-solid interface, under the condition of open-pore condition, that is, when the porous fluid of Composition 1 and Composition 2 in the medium are connected, it is easy to obtain the following boundary conditions [48,76]

$$\begin{cases} T_i^{(1)} = T_i^{(2)}, \\ \sigma^{(1)} = \sigma^{(2)}, \end{cases} \quad (43b)$$

and

$$\begin{cases} u_i^{(1)} = u_i^{(2)}, \\ w_n^{(1)} = w_n^{(2)}. \end{cases} \quad (43c)$$

In Eq. (43c), $n$ stands for the normal direction of the interface. Eqs. (43b) and (43c) can also be extended to multiphase porous media. If there is no relative motion, then the boundary conditions are consistent with the case of solid-solid media by considering all the other displacements to be continuous.

So far, based on energy conservation and mass conservation, the boundary conditions for the stress and displacement on an interface are formulated in porous media under the open pore conditions. Similarly, the treatment for boundary condition determinations can be easily extended to any combination of multiphase porous media. It allows us to use energy and mass conservations study the wave propagation in arbitrary multiphase porous media and the reflection and refraction at various interfaces. The deduced continuation conditions are consistent with the existing ones.

## 5. Conclusions and discussions

In continuum mechanics, energy conservation equation, i.e., the rate of total mechanical energy with time equals to the rate of the work with time done by external forces is spelt out as an axiom. From this axiom, Lagrange's equation, Hamilton's equations, and the elastodynamic equation of motion for arbitrarily anisotropic and multiphasic poroelastic media are formulated for the first time, which are consistent with the equation formulations using Hamilton's principle, also for the first time. The boundary conditions for various elastic media, including porous ones are directly derived and extended from the principle of energy conservation as an application example of our methodology.

The advantages of our methodology are that it does not have to introduce the concepts of variational principles, neither is there a need for time terminal constraints. It is simple and easy to understand, and its physical meaning is clear, especially suitable for the formulations of weak form dynamic equations in multiphysics coupling systems.

It is pointed out that in continuum mechanics, contrary to the existing Hamiltonian mechanics, Hamilton's principle, Lagrange's equation, and Newton's second law of motion, are all the consequences of the principle of energy conservation. The energy conservation not only can be used to explain the behaviors of the time-space evolution of wave motion in a broad sense, but also it can be used to elaborate the wave motion mechanisms on complex boundaries, such as wave reflections and refractions at interfaces, i.e., Snell's laws of acoustic wave reflection and refraction are also the consequence of the energy conservation.

Our proposed methodology unlocks the essences of Hamilton's principle in classical mechanics, and endows Hamilton's principle a new physical explanation, i.e., the reason why the true motion trajectories of the particle points predicted by Hamilton's principle, is energy conservation, and essentially it is energy conservation that governs the nature's motion. Our methodology may be extended to the other areas for equation formulations, such as electrodynamics, fluid mechanics, etc.

## Acknowledgments

This work is supported by the National Natural Science Foundation of China with Grant Nos. 11734017, 11974018, 12274432 and 52227901, respectively.This work is supported by the National Natural Science Foundation of China with Grant Nos. 11734017, 11974018, 12274432 and 52227901, respectively.